\documentclass[
aps,prd,
showpacs,twocolumn,notitlepage,
amssymb,amsmath,amsfonts,mathrsfs,
nofootinbib,superscriptaddress,
floats,floatfix
 amsmath,amssymb,
 aps,
]{revtex4-2}
\usepackage{aas_macros}
\usepackage{graphicx}
\usepackage{xcolor}
\usepackage[colorlinks=true]{hyperref}
\hypersetup{citecolor=cyan,linkcolor=magenta}
\usepackage{multirow,array}

\usepackage{amsmath}
\usepackage{graphics}
\usepackage{epstopdf}
\usepackage{enumerate}

\usepackage{savesym}
\savesymbol{tablenum}
\usepackage{siunitx}
\restoresymbol{SIX}{tablenum}
\newcommand{\eeq}{\end{equation}}
\newcommand{\beqn}{\begin{eqnarray}}
\newcommand{\eeqn}{\end{eqnarray}}
\newcommand{\pa}{\partial}

\newcommand{\cR}{{\cal{R}}}

\newcommand{\cC}{{\cal{C}}}

\usepackage{siunitx}
\usepackage{graphicx}
\usepackage{dcolumn}
\usepackage{bm}

\begin{document}

\title[]{Resonances of compressible stars in precessing orbits around a spinning black hole}

\author{Matteo Stockinger}

\affiliation{Max-Planck-Institut f\"ur Gravitationsphysik (Albert-Einstein-Institut), Am M\"uhlenberg 1, D-14476 Potsdam-Golm, Germany}
\email{matteo.stockinger@aei.mpg.de} 

\author{Masaru Shibata}
\affiliation{Max-Planck-Institut f\"ur Gravitationsphysik (Albert-Einstein-Institut), Am M\"uhlenberg 1, D-14476 Potsdam-Golm, Germany}
\affiliation{Center for Gravitational Physics and Quantum Information, Yukawa Institute for Theoretical Physics, Kyoto University, Kyoto, 606-8502, Japan}

\begin{abstract}
In our previous paper, we reported the presence of a new resonance of an incompressible star orbiting a spinning black hole and showed that it can set in before the tidal disruption limit 
if the star has an inclined spherical orbit around the black hole. 
Using the affine model developed by Carter and Luminet, we extend our result to the stars with polytropic equations of state. We give further credence to the result previously found. We also derive the formula for the growth rate of the resonant motion, which is useful for checking the results of hydrodynamics simulations.
\end{abstract}

\maketitle

\section{Introduction}


Tides play an important role in numerous astrophysical phenomena, 
ranging from tidal disruption events to the dynamics of binary objects~\cite{ogil14, geza21, souc13}. As such they have been long studied~\cite{cart99, depa13}, using both analytical and numerical methods. 

Tides is generally studied assuming Newtonian gravity. Such studies have thus elucidated valuable effects of tides in binary systems involving planets and main-sequence stars, in particular on their orbits~\cite{bode01, ibgu09, ogil14}. However in order to understand the impact of tides on binary compact objects, relativistic effects have to be taken into account. Many articles have explored how a relativistic framework modifies the tidal effects. For example, tides are known to have an appreciable impact on the late inspiral of binary compact stars~\cite{koch92, bild92, 1993ApJ...406L..63L, 1994ApJ...420..811L, 2008PhRvD..77b1502F}. The condition for the tidal disruption is known to be modified appreciably by general relativistic effects~\cite{Rel_RocheI, Rel_RocheII, shib96, wigg00, Ishii:2005xq}. 

However, until quite recently, there have been few studies delving into the case of a star orbiting a Kerr black hole (BH) in a non-equatorial orbit~\cite{bane19}, for which there is an orbital precession with respect to the equatorial plane. This orbital precession can be also induced for a star orbiting a highly deformed companion with a quadrupole moment. To fully understand the effects of tides in general binary systems, it is necessary to investigate the impact of orbital precession. Using the virial equations to model the star as an incompressible fluid~\cite{chan69}, we studied the influence of the tidal forces of the BH on stars in a precessing spherical orbit with a small orbital separation~\cite{stoc24} and found that resonances could arise in such systems, giving a further example of tidal resonances in astrophysics.

Tidal resonances are a particular class of tidal phenomena which can play an important role in many astrophysical systems, from planets to neutron stars. They typically happen when the tidal forces acting on an object excites one of its normal modes. Tidal resonances have been encountered for f-modes, 
g-modes, 
as well as for inertial modes (relying on the Coriolis force)~\cite{Shibata:1993qc, lai94, reis94, kokk95, flan07, xu17, pois20, asto23, yu24}. They could modify the orbital evolution of binary systems. This indicates that understanding their effects are paramount for a better understanding of the dynamics of such systems. 

Our previous study found the tidal resonance for stars in precessing spherical orbits around Kerr BHs assuming an incompressible fluid with which all the analyses can be analytically carried out~\cite{stoc24}. In this paper, we do not assume this anymore. To enable us to derive important formula analytically, we use the affine star model \cite{cart85} assuming a polytropic equation of state (EOS). This model has already been widely used to study particular tidal problems analytically but still quite precisely~\cite{cart83, lumi85}. In \cite{cart85}, the authors also prove a conformal correspondence theorem, which links the results found using the virial equations, with more realistic models of stars. In this paper, we use this theorem to extend our previous result, and figure out whether we can see the previously-found tidal resonances in stars modeled with polytropic EOS. 

As in \cite{stoc24}, we carry out a perturbative analysis for a star of precessing spherical orbits using the result found for the star with a circular orbit in the equatorial plane as the zeroth-order solution. We show the presence of the same resonances if the EOS is not extremely soft. This stems from the similarity between the equations used in incompressible and affine models. We also show that as the EOS becomes softer, the resonant distance diminishes.

This paper is organized as follows: In Sec.~\ref{affineModel}, we review the affine model, and show how it describes our system. In Sec.~\ref{Per}, we present our method for a perturbative expansion with respect to the inclination angle, which we apply up to the second order. We also derive a formula for the growth rate of the resonant motion. In Sec.~\ref{Res}, we compute the new resonant distances of our system focusing on white dwarfs orbiting massive BHs, and show how it changes for polytropic EOSs with different polytropic indices. We also indicate that in the presence of the resonance, mass ejection is likely to be induced from the stellar surface. 
Throughout this paper, we use the geometrical units of $c=G=1$, where $c$ and $G$ are the speed of light and gravitational constant, respectively, and the Einstein summation convention.

\section{Affine model}

\label{affineModel}

\subsection{Setup}

In our previous paper \cite{stoc24}, we modeled the orbiting star as an incompressible ellipsoidal object on which the tidal forces by the BH are exerted. Newtonian gravity was assumed for the self-gravity part of the star. 
To go beyond the incompressible hypothesis, we employ the affine model, which was originally defined by Carter and Luminet~\cite{cart83,cart85}. 

The affine model was introduced in \cite{cart83} to analytically investigate stars which are tidally deformed by massive BHs. This model was further developed in~\cite{cart85}, which allowed the authors to show that most of the properties of the incompressible model could be carried over to stars with more general EOSs. Our present purpose is to investigate the presence of the resonance found in our previous paper~\cite{stoc24} changing the EOS from the incompressible one to a softer one (changing the polytropic index from $n=0$ to $<3$). The affine model enables us to extend the aforementioned resonance to compressible EOSs. 

The affine star model assumes that the position vector $r_i$ in the center or mass frame of any material element is given by \begin{eqnarray}
\label{eq1}
    r_i=q_{ia}\hat r_a,
\end{eqnarray} 
where indices from the middle of the Latin alphabet $i,j,k,...$ refer to spatial components in the physical frame, while letters from the beginning of the alphabet $a,b,c,...$ refer to the spherical reference state. Whenever the same subscript appears like $q_{ia}\hat r_a$, we assume to take the sum for the three spatial components. 

$q_{ia}$ is the deformation matrix and is assumed to be spatially uniform. The spherical reference state is the state that the star should have if it is static and completely isolated: In this case, the self-gravity of the fluid should enforce the star to be spherical.

Since we assume that $q_{ia}$ is spatially uniform, the affine model applies a linear transformation to the reference state. Hence the kinematics of the star is completely contained within the nine components of the deformation matrix $q_{ia}$.
Furthermore, as the deformations are assumed to be linear, the star has an ellipsoidal shape, and the surfaces of constant density are self-similar ellipsoids as in the incompressible case~\cite{chan69}. 

This idealized affine model can also be derived as approximate equations of the full hydrodynamics equations. This can be seen by expanding the relative position vector $r_i$ in a Taylor series of $\hat r_a$ \begin{eqnarray}
    r_i=q_{ia}\hat r_a+q_{iab}\hat r_a\hat r_b+O(|\hat r|^3).
\end{eqnarray} 
From this perspective, it can be seen that neglected terms in the Taylor expansion will be in general important in the outer part of the star (unless $q_{iab}$ is extremely large), and thus only for a small portion of the total mass. Overall the affine model can be seen as a good approximation when considering linear forces acting on a self-gravitating object. As we assume only linear tidal forces acting on the star, we may expect that the results found using this model should be robust. 

This model was shown to be suitable for exploring tidal deformation of stars with moderate degrees of penetration within the tidal disruption radius \cite{cart83, lumi85, lumi86}. However, when the penetration is too deep, the non-linear deformation of the star becomes so important that the affine model breaks down. We can also expect it to fail in other scenarios. For example, if the star gets quite close to the black hole, we have to consider non-linear contributions to the tidal tensor~\cite{Ishii:2005xq}. Another case would be when the star gets past the Roche limit, i.e. when stellar matter gets ejected from the star. In this paper we do not consider these cases.

We neglect the general relativistic effect for the self-gravity of the star, so there could be an error of order $Gm_{\text{star}}/c^2 R_{\text{star}}$. However, for typical-mass white dwarfs with $m_\mathrm{star}=0.6$--$0.7M_\odot$ which we consider in this paper, such a correction is minor. 

As we already mentioned, we assume that the EOS of the star is polytropic,  $P=\kappa\rho^{\Gamma}=\kappa\rho^{1+1/n}$, with $P$ the pressure, $\rho$ the rest-mass density, $\Gamma=1+1/n$, $n$ the polytropic index, and $\kappa$ the polytropic constant. The spherical reference state is then computed by solving the Lame-Emden equation~\cite{1939isss.book.....C}
\begin{eqnarray}
    \frac{1}{\xi^2}\frac{d}{d\xi}\left(\xi^2\frac{d\theta}{d\xi}\right)+\theta^n=0,
\end{eqnarray} 
where $\xi$ and $\theta$ are related to the radial coordinate $r$ and the density of the spherical reference state $\hat\rho$ by
\begin{eqnarray}
    r&=&\sqrt{\frac{(n+1)\kappa\rho_c^{1/n-1}}{4\pi}}\xi,\\
    \hat\rho&=&\rho_c\theta^n,
\end{eqnarray} and $\rho_c$ is the central density of the star. $\theta$ is unity at the stellar center and zero at the surface. 

The hydrodynamics equation for self-gravitating stars subject to tidal forces in an inertial frame is written as
\begin{eqnarray}
    \rho\ddot r_i=-\rho\partial_i\Phi-\partial_iP-\rho C_{ij}r_j,
\end{eqnarray} 
where 
$\Phi$ is the self-gravitational potential, determined by the Poisson equation, $\Delta\Phi =4\pi \rho$, and $C_{ij}$ is the tidal tensor determined by a BH, which is appropriately defined in the local inertial frame~\cite{marc83}. $\dot r_i$ denotes the derivative of $r_i$ with respect to the timelike affine parameter that characterizes the motion of the inertial frame. 

The first step to obtain the basic equations satisfied by the affine model is to recover the tensor virial equation. For this, we take the tensor product of the preceding hydrodynamics equation with the position vector and then integrate over the volume of the star $V$ as
\begin{eqnarray}
    \int_V\rho(r)\ddot r_ir_kd^3r=&&-\int_V\rho(r)\partial_i\Phi(r) r_kd^3r-\int_V\partial_iP(r)r_kd^3r\nonumber\\
    &&-\int_V\rho(r) C_{ij}r_jr_kd^3x.
\end{eqnarray}
By introducing the following integral quantities \begin{eqnarray}
    \mathcal{M}_{ik}&=&\int_V\rho(r) r_ir_kd^3r,\\
    J_{ik}&=&\frac{1}{2}\int_V\rho(r)(r_i\dot r_k-\dot r_ir_k)d^3r,\\
    T_{ik}&=&\frac{1}{2}\int_V\rho(r) \dot r_i\dot r_kd^3r,\\
    \Omega_{ik}&=&\int_V\rho(r) \partial_i\Phi \, r_kd^3r,\\
    \Pi&=&\int_VPd^3r, 
\end{eqnarray} 
we obtain the tensor virial equation~\cite{1954PhRv...96.1686P}
\begin{equation}
    \frac{1}{2}\ddot{\mathcal{M}}_{ik}+\dot J_{ik}-2T_{ik}=\Omega_{ik}+\Pi\delta_{ik}-C_{ij}\mathcal{M}_{jk},
\end{equation} where $\delta_{ik}$ is the Kronecker delta.

In the spherical reference state, we have \begin{eqnarray}
    \int_V\hat\rho(r)\hat r_a\hat r_bd^3\hat r=\mathcal{M}_*\delta_{ab},
\end{eqnarray} where \begin{eqnarray}
    \mathcal{M}_*=\frac{1}{3}\int_V\hat\rho(\hat r)\hat r^2d^3\hat r.
\end{eqnarray}
As we apply the linear transformation $q_{ia}$, the mass of an infinitesimal material element is conserved, and hence, we have the relation $\hat\rho(\hat r)d^3\hat r=\rho(r)d^3 r$. We then obtain the following relations for the integral quantities: 
\begin{eqnarray}
    \mathcal{M}_{ik}&=&\mathcal{M}_*q_{ia}q_{ka},\\
    J_{ik}&=&\frac{\mathcal{M}_*}{2}(q_{ia}\dot q_{ka}-\dot q_{ia}q_{ka}),\\
    T_{ik}&=&\frac{\mathcal{M}_*}{2}\dot q_{ia}\dot q_{ka},
\end{eqnarray} 
and thus, 
\begin{eqnarray}
    \mathcal{M}_*\ddot q_{ia}=q^{-1}_{ak}(\Omega_{ik}+\Pi\delta_{ik})-C_{ij}\mathcal{M}_*q_{ja}.
\end{eqnarray}

If we introduce $B_{ij}$ by
\begin{eqnarray}
    \mathcal{M}_*B_{ij}=q^{-1}_{ai}q^{-1}_{ak}(\Pi\delta_{kj}+\Omega_{kj}),
\end{eqnarray} 
the equation for $q_{ia}$ becomes \begin{eqnarray}
    \ddot q_{ia}=(B_{ij}-C_{ij})q_{ja}. 
\end{eqnarray}
$B_{ij}$ thus encodes the internal forces of the star, namely the pressure and the self-gravity. In particular, if we set $C_{ij}$ to zero, the resulting equation describes the evolution for an isolated star.
As we assume that the self-gravity is Newtonian, we have \begin{eqnarray}
    \Omega_{ij}=-\int_V\int_V\frac{(r_i-r'_i)(r_j-r'_j)\rho(r)\rho(r')d^3rd^3r'}{2|r_k-r'_k|^3}.~~
\end{eqnarray} 

We then define the matrix $\mathbf{S}$, as $S_{ij}=q_{ia}q_{ja}$. By denoting the identity matrix as $\mathbf{1}$, we introduce the dimensionless matrix $A_{ij}$ 
\begin{eqnarray}
A_{ij}=\sqrt{\det(\mathbf{S})}\int_0^{\infty}\frac{(\mathbf{S}+u\mathbf{1})_{ij}^{-1}du}{\sqrt{\det(\mathbf{S}+u\mathbf{1})}}. \label{eq22}
\end{eqnarray} 
From the theory of incompressible fluids \cite{chan69, cart85}, we have \begin{eqnarray}
    \Omega_{ij}=\frac{\Omega_*}{2\sqrt{\det(\mathbf{S})}}A_{ik}S_{jk},
\end{eqnarray} where $\Omega_*$ is the self-gravitational energy in the spherical reference state given by
\begin{eqnarray}
    \Omega_*=-\int_{V_*}\int_{V_*}\frac{\rho(\tilde r)\rho(\tilde r')d^3\tilde rd^3\tilde r'}{2|\tilde r_k-\tilde r'_k|},
\end{eqnarray} 
and $V_*$ designates the volume of the star in the spherical reference state. 

If we also introduce the integral of the pressure in the spherical reference state $\Pi_*$ by
\begin{eqnarray}
    \Pi_*=\int_{V_*}P(\tilde r)d^3\tilde r,
\end{eqnarray} 
we must have \begin{eqnarray}
    \Pi_*=-\frac{\Omega_*}{3}.
\end{eqnarray} 
This is equivalent to the virial relation for isolated spherical stars.
We can now rewrite $B_{ij}$ as 
\begin{eqnarray}
    B_{ij}=\frac{\Pi_*}{\mathcal{M}_*}\left(-\frac{3}{2\sqrt{\det(\mathbf{S})}}A_{ij}+\frac{\Pi}{\Pi_*}q^{-1}_{ai}q^{-1}_{aj}\right).
\end{eqnarray}
Then the equation of $q_{ia}$ is written as
\begin{eqnarray} \label{eqInertFrame}
    \ddot q_{ia}=\left(\frac{\Pi_*}{\mathcal{M}_*}\left(-\frac{3}{2\sqrt{\det(\mathbf{S})}}A_{ij}+\frac{\Pi}{\Pi_*}q^{-1}_{bi}q^{-1}_{bj}\right)-C_{ij}\right)q_{ja}. \nonumber \\
\end{eqnarray}

\subsection{Equations in a rotating frame}

In the affine model, we assume that all the degrees of freedom of the system is contained within the deformation matrix $q_{ia}$. Thus in the most general case, we have nine free variables. 

For any $q_{ia}$ that satisfies det$(q_{ia})\not=0$, we can find two independent orthogonal matrices $P_{ij}$ and $Q_{ab}$, i.e., $P_{ij}=P^{-1}_{ji}$ and $Q_{ij}=Q^{-1}_{ji}$, and write
\begin{eqnarray}
    q_{ib}=P_{ij}b_{ja}Q_{ba}.
\end{eqnarray} 
This defines $b_{ja}$ as a new deformation matrix written in a new frame. This frame is defined through a rotation in real space $P_{ij}$, and a rotation in the spherical reference state $Q_{ab}$.
Since we do not impose any constraint on $b_{ia}$ at this stage, it has the same nine degrees of freedom as $q_{ia}$.

To relate quantities between the inertial frame and our new frame, we define the rotation rate of the inertial frame \begin{eqnarray}
    w_{ij}=P_{ki}\dot P_{kj}.
\end{eqnarray} 
Likewise we define the vorticity inside the star through \begin{eqnarray}
    \lambda_{ab}=Q_{ca}\dot Q_{cb}.
\end{eqnarray}
Both of them are anti-symmetric matrices, i.e., $w_{ij}=-w_{ji}$ and $\lambda_{ab}=-\lambda_{ba}$, and thus, each of them has three degrees of freedom.



We then have the relations 
\begin{eqnarray}
    P_{ji}\ddot q_{jb}Q_{ba}=&&\ddot b_{ia}+2w_{ij}\dot b_{ja}-2\dot b_{ib}\lambda_{ba}+\dot w_{ij}b_{ja}-b_{ib}\dot\lambda_{ba}\nonumber\\
    &&+w_{ij}w_{jk}b_{ka}-2w_{ij}b_{jb}\lambda_{ba}+b_{ic}\lambda_{cb}\lambda_{ba}.
\end{eqnarray}
Thus the equation of the affine model becomes 
\begin{eqnarray}
    \ddot b_{ia}&&+2w_{ij}\dot b_{ja}-2\dot b_{ib}\lambda_{ba}+\dot w_{ij}b_{ja}-b_{ib}\dot\lambda_{ba}+w_{ij}w_{jk}b_{ka}\nonumber\\
    &&-2w_{ij}b_{jb}\lambda_{ba}+b_{ic}\lambda_{cb}\lambda_{ba} \nonumber \\
    &&-\tau_*^{-2}\left(\frac{\Pi}{\Pi_*}b^{-1}_{ia}-\frac{3}{2\sqrt{\det(\mathbf{\tilde S})}}\tilde A_{ij}b_{ja}\right)
    =-\tilde C_{ij}b_{ja}, ~~\label{eq32}
\end{eqnarray} 
where $\tau_*^{-2}=\Pi_*/\mathcal{M}_*$, $\tilde S_{ij}=b_{ia}b_{ja}$, $\tilde C_{ij}$ is the tidal tensor $C_{ij}$ in the new frame, and $\tilde A_{ij}$ has the same expression as $A_{ij}$ but with $S_{ij}$ replaced by $\tilde S_{ij}$ in Eq.~\eqref{eq22}. The new equation \eqref{eq32} explicitly shows the contribution of rotation and vorticity on the star through the terms containing $w_{ij}$ and $\lambda_{ab}$. $\tau_*$ approximately denotes the sound crossing time in the spherical reference star.

If we choose a frame in which $b_{ia}$ is diagonal, i.e. $b_{ia}=b_i\delta_{ia}$,  this promotes $w_{ij}$ and $\lambda_{ab}$ to dynamical variables which always enforce the relation $q_{ib}=P_{ij}b_{ja}Q_{ba}$. We still have nine degrees of freedom, but in this case each of $b_i$, $w_{ij}$ and $\lambda_{ab}$ has three degrees of freedom.

We introduce the rotation vector $\Omega_k$ and the vorticity vector $\Lambda_c$ such that $w_{ij}=\epsilon_{ijk}\Omega_k$ and $\lambda_{ab}=\epsilon_{abc}\Lambda_c$ where $\epsilon_{ijk}$ is the completely antisymmetric tensor. We can then decompose Eq.~\eqref{eq32} as 
\begin{eqnarray}
&&    \ddot b_i-b_i[(\Lambda_j^2+\Lambda_k^2)+(\Omega_j^2+\Omega_k^2)]+2(b_j\Lambda_k\Omega_k+b_k\Lambda_j\Omega_j)
    \nonumber \\
  &&  ~~~~-\tau_*^{-2}\left(\frac{\Pi}{\Pi_*b_i}-\frac{3}{2b_1b_2b_3}b_i\tilde A_{ii}\right)=-\tilde C_{ii}b_i, \label{eq33}\\
  &&
    2 \dot b_i\Lambda_k-2\dot b_j\Omega_k+b_i\dot\Lambda_k-b_j\dot\Omega_k+b_i\Lambda_i\Lambda_j+b_j\Omega_j\Omega_i
    \nonumber \\
    && ~~~~-2b_k\Lambda_i\Omega_j=-\tilde C_{ij}b_j, \label{eq34}
\end{eqnarray} 
where we suspended the summation convention, and assumed $i\neq j\neq k$. Using $b_i$, $\tilde A_{ij}$ is written as
\begin{eqnarray}
\tilde A_{ij}&=&b_1b_2b_3\int_0^{\infty}\frac{\delta_{ij}du}{\sqrt{(b_1^2+u)(b_2^2+u)(b_3^2+u)}(b_i^2+u)}. \nonumber \\
\end{eqnarray} 
It is found that $\tilde A_{ij}$ is written in a form well-known in the incompressible case~\cite{chan69}. Equations~\eqref{eq33} and \eqref{eq34} correspond to the evolution equations for the diagonal and off-diagonal components of Eq.~\eqref{eq32}, respectively.
Thus rewritten, we have similar equations to the ones of the Dirichlet problem as shown in~\cite{chan69}.




\subsection{Equatorial plane - Order 0}
\label{equatPlane}

Our purpose in this paper is to study the tidal effect on a polytropic star orbiting a Kerr BH using the affine model. Apart from using a compressible EOS, our setup completely mirrors the one presented in \cite{stoc24}.

\begin{figure}[t]
    \centering
        \includegraphics[width=0.5\textwidth]{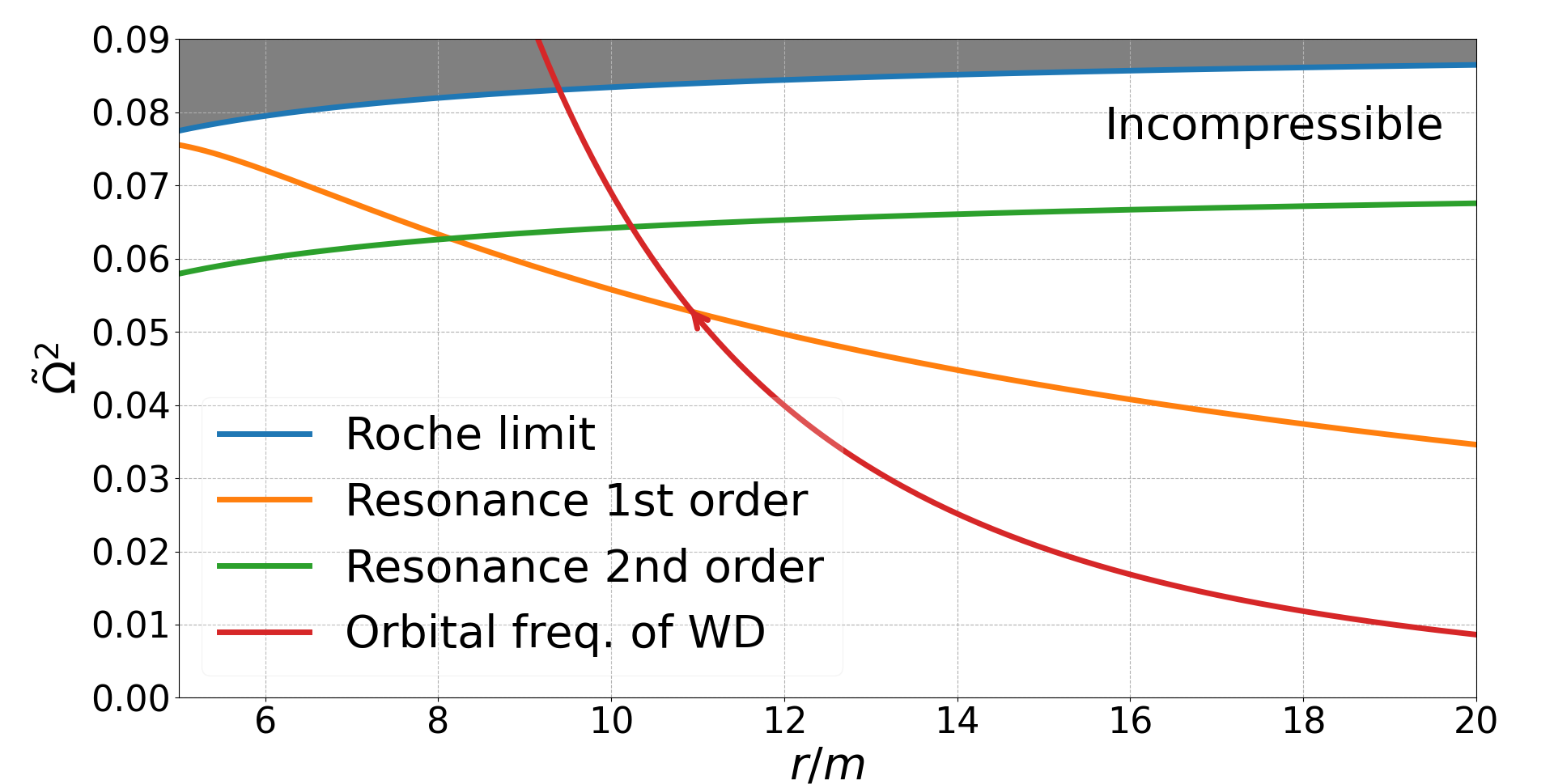}
    \hfill
         \includegraphics[width=0.5\textwidth]{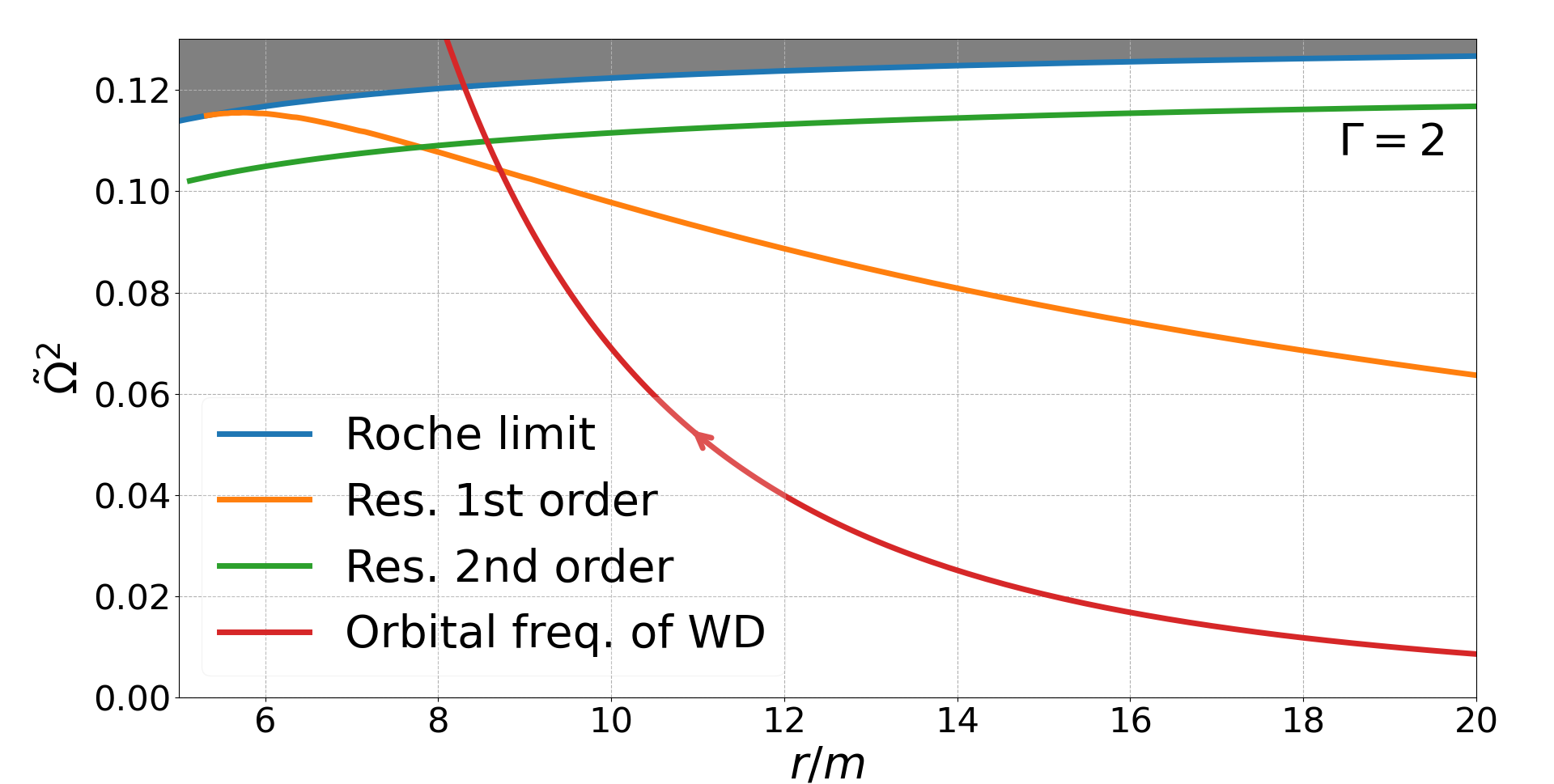}
    \caption{\label{Reson}$\tilde\Omega^2=\Omega^2/(\pi\overline\rho)$ as a function of $r/m$ for the first- and second-order resonances, along with the tidal disruption limit $\tilde\Omega^2_{\mathrm{crit}}$ as well as the orbital frequency of $\tilde\Omega^2$ for a typical value of supermassive BH-white dwarf binary, namely a $10^4$ km large, $0.6 M_{\odot}$ white dwarf orbiting a $10^5 M_{\odot}$ BH (red curve with arrow). The orange and green curves denote the location of the first- and second-order resonances, respectively (see Sec.~\ref{Res} for details). For all the plots, we assumed $a/m=0.8$ and prograde orbits. The upper panel shows the results for the incompressible model and the lower panel the result for compressible stars with $n=1$ ($\Gamma=2$). 
    }
\end{figure}

\begin{figure}[t]
    \centering
         \includegraphics[width=0.5\textwidth]{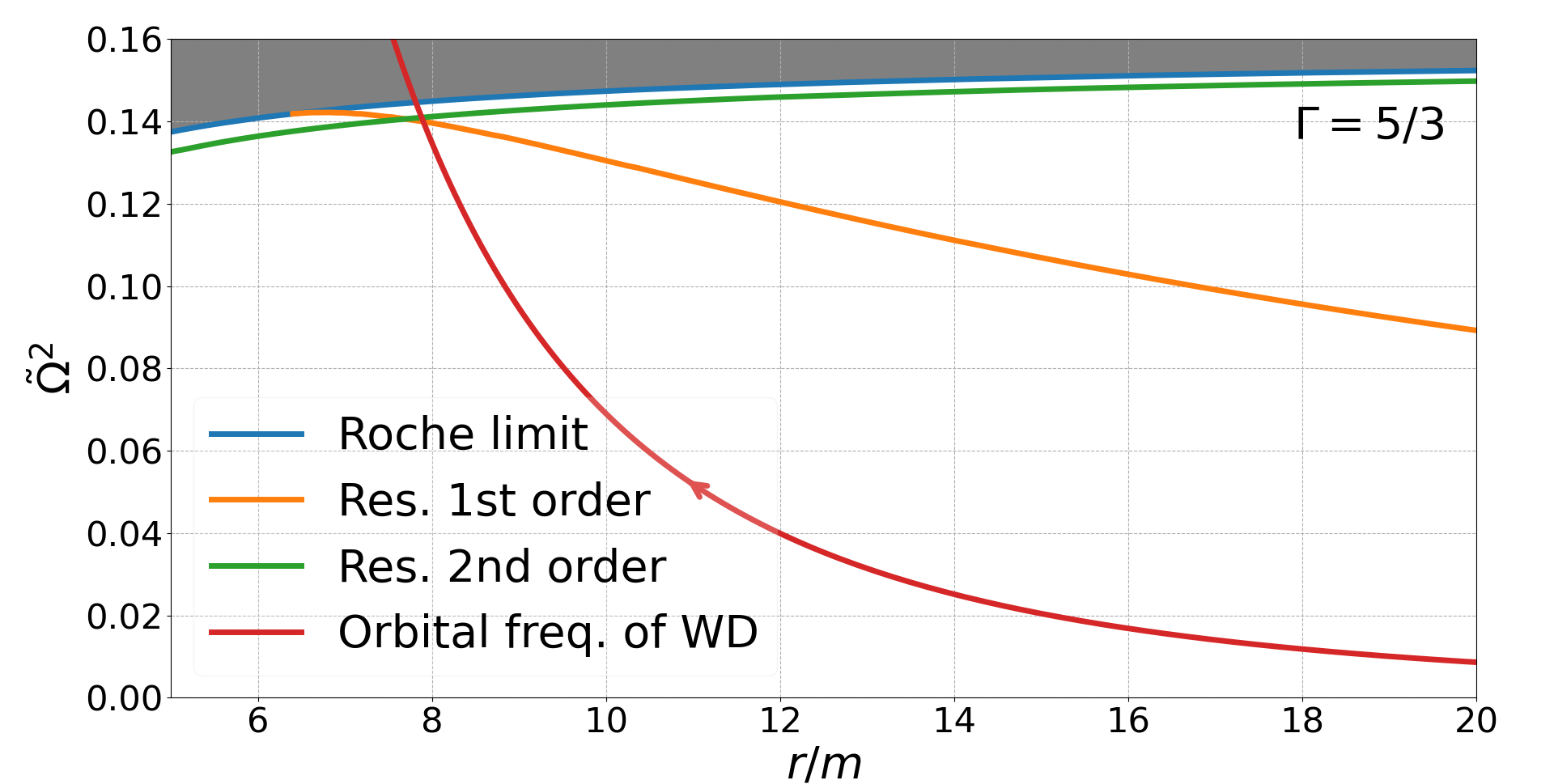}
    \hfill
         \includegraphics[width=0.5\textwidth]{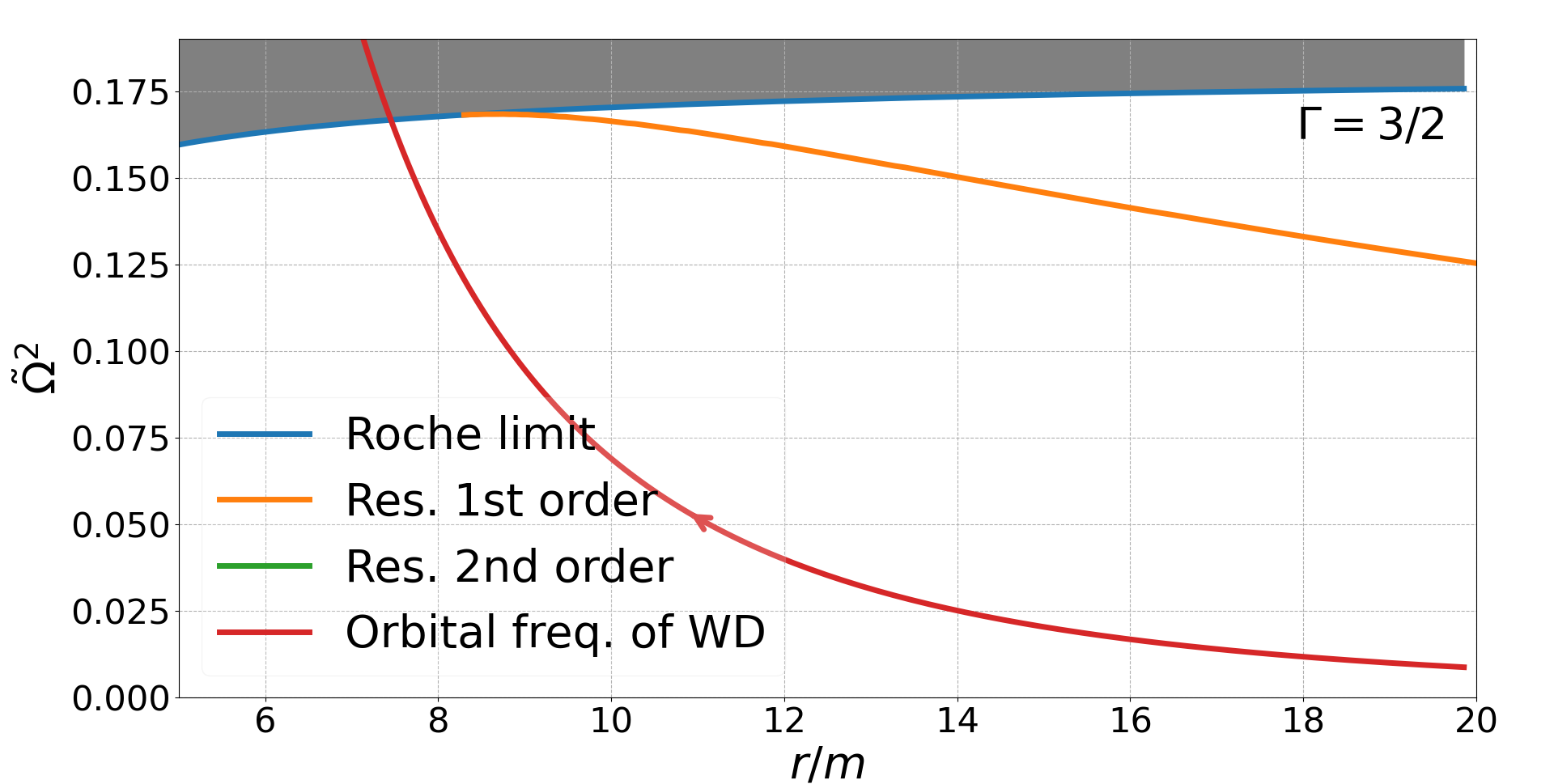}
    \caption{\label{Reson2}Same curves as Fig.~\ref{Reson}, but the upper and lower panels show the results for compressible stars with $n=3/2$ and 2 ($\Gamma=5/3$ and $3/2$), respectively.
    }
\end{figure}

As such, we assume that the star has a geodesic orbit around a spinning BH of mass $m$ and of spin  parameter $a$. To reference the position of the star relative to the BH, we will use the Boyer-Lindquist coordinates. The metric and the geodesic equations are written down in Appendix \ref{geodesic}. We will assume spherical orbits of the star; the star stays at constant radius $r$ of the BH but its orbital plane slightly precesses around the equatorial plane.
We assume that the mass ratio between the two objects as $m_{\text{star}}/m\ll1$ (typically $10^{-5}$ in this paper). Thus, we can safely treat the spacetime as fixed because the modification of the main results by changing the equations of state (see Figs.~\ref{Reson} and \ref{Reson2}) is much more significant than the correction from the mass ratio. 

The equation of the affine model (\ref{eqInertFrame}) are valid in a local inertial frame. Thus, we need to use Fermi-normal coordinates on the geodesics, such as the ones found in~\cite{marc83}. 

In this inertial frame, we refer to the axes of the corresponding orthonormal basis as 1, 2, and 3. We assume that the axis 3 points in the same direction as the orbital angular momentum of the fluid or in other words, the direction of $-(\pa/\pa\theta)^\mu$. Note that our 2 and 3 axes agree with the 3 and 2 axes of \cite{marc83}, and that we changed the direction of our 2-axis. In this case, the tidal tensor has the form given in Appendix \ref{tidalTens}.


As the zeroth order solution, we need to obtain the solution of a star with an equatorial orbit. 
As in \cite{shib96}, we assume that the semi-axes of our ellipsoidal star point in the frame rotating at the orbital frequency and put ourselves in this frame. We thus use as ansatz \begin{eqnarray}
    b_{ja}&=&b_a\delta_{ja}, \\
    w_{ij}&=&w\epsilon_{ij3},
\end{eqnarray} 
where $w$ is the orbital angular velocity which satisfies $w^2=m/r^3$. For equatorial orbits, we then have $\dot w_{ij}=0$. In this paper, we pay attention only to the case that the star has no vorticity; thus $\lambda_{ab}=0$, i.e., the corotating case.

We can assume that $b_{ia}$ is constant in time for the equatorial orbits case. Then, Eq.~\eqref{eq32} reduces to 
\begin{eqnarray}
    w_{ij}w_{jk}b_{ka}-\tau_*^{-2}\left(\frac{\Pi}{\Pi_*}b^{-1}_{ia}-\frac{3}{2b_1b_2b_3}\tilde A_{ij}b_{ja}\right)=\tilde C_{ij}b_{ja}. \nonumber \\
\end{eqnarray}

To utilize the results previously found~\cite{stoc24}, we introduce the conformal correspondence theorem \cite{cart85}. This theorem relates the result of the equation previously written to the results found in the incompressible hypothesis. For this, we first rescale the variables as
\begin{eqnarray}
    \alpha&=&(b_1b_2b_3)^{1/3},\\
    b_{ia}&=&\alpha\hat b_{ia},\\
    w_{ij}&=&\alpha^{-3/2}\tau_*^{-1}\hat w_{ij}.
\end{eqnarray}
We then obtain \begin{eqnarray}
\label{confCorrTh}
    \hat w_{ij}\hat w_{jk}\hat b_{ka}-\left(\hat P\hat b^{-1}_{ia}-\frac{3}{2}\tilde A_{ij}\hat b_{ja}\right)=\tilde C_{ij}\hat b_{ja},
\end{eqnarray} 
where $\alpha\Pi/\Pi_*=\hat P$. In a polytropic EOS~\cite{cart85}, we have $\alpha=\hat P^{1/(4-3\Gamma)}$. Due to this expression, we cannot extend this formalism to the particular case $\Gamma=4/3$.


To solve the equation verified by a compressible star, we thus need to find a solution for the virial equations of an incompressible one. By solving these equations, we get the values of $\hat b_1, \hat b_2$, and $\hat b_3$. This allows us to compute $\hat P$ which gives us $\alpha$. We can then rescale the star appropriately.

It is also worth noting how rescaling affects the density of the compressible star. If the spherical reference state of our compressible star has a central density of $\rho_{c,*}$ in its physical state, it will have a density of $\rho_c=\alpha^{-3}\rho_{c,*}$.

However if we want to actually find the right physical values, we need to proceed carefully. Indeed when we rescale the angular velocity $w$, we change the separation of the star to the BH, $r$. As in the incompressible case, we take $w^2=m/r^3$, and thus, by rescaling $w$, we rescale $r$ as $r=\alpha\tau_*^{2/3}\hat r$. 

Here, in our equation, $\tilde C_{ij}$ depends on $r$. In particular, we can write it as $\tilde C_{ij}=w^2\tilde c_{ij}(r)$. When rescaling, we have $\hat C_{ij}=\hat w^2\tilde c_{ij}(\alpha\tau_*^{2/3}\hat r)$. Thus we cannot fully remove all $\alpha$ and $\tau_*$ from the rescaled equations. More specifically we cannot compute all the equations all at once. 
Therefore, we can solve the equations only by using an iterative scheme.

At the beginning of the iteration, we fix $\tilde c_{ij}(r)$ using the actual separation $r$ between the star and the BH. Furthermore $\tau_*$ is fixed from the spherical reference state of the star for which we want to solve the equations.
Then, we assume $\alpha=1$. This fixes $\hat w$, and we can then solve Eq.~\eqref{confCorrTh}, in the same way as in \cite{shib96}. Indeed as we assumed that the only non-zero components of $b_{ia}$ are diagonal, we have the equations 
\begin{eqnarray}
    0&=&-\frac{3}{2}\hat b_1\tilde A_{11}+\frac{\hat P}{\hat b_1}-\hat b_1\tilde C_{11}-\hat w^2\hat b_1,\\
    0&=&-\frac{3}{2}\hat b_2\tilde A_{22}+\frac{\hat P}{\hat b_2}-\hat b_2\tilde C_{22}-\hat w^2\hat b_2,\\
    0&=&-\frac{3}{2}\hat b_3\tilde A_{33}+\frac{\hat P}{\hat b_3}-\hat b_3\tilde C_{33}.
\end{eqnarray}
From these equations, we can compute $\hat P$, which yields a new value for $\alpha_{\text{new}}=\hat P^{1/(4-3\Gamma)}$. In general this new value will be different from the one used previously.

Thus we change the value for $\alpha$ accordingly, and this changes the value for $\hat w$. We then solve again the equations. We continue this process until there is a good agreement between the value of $\alpha$ used to solve the equations and the one we get from $\hat P$.

As in previous works~\cite{Rel_RocheI, shib96, wigg00}, we find a distance below which there is no more possible solution. This will correspond to the Roche limit. We plot the maximum angular velocity $\tilde\Omega^2=\Omega^2/(\pi\overline\rho)$ as a function of the separation between the star and the BH in Figs.~\ref{Reson} and \ref{Reson2} (line delimiting the shaded region from the non-shaded one), where $\overline\rho$ is the average density of the star. If a star is more compact than another one, then its value for $\tilde\Omega$ will be lower. Thus it will reach the Roche limit at a closer distance.

Figure~\ref{Reson} shows the results for stiff EOSs, i.e., the incompressible EOS and polytropic EOS with $\Gamma=2$ and Fig.~\ref{Reson2} for the soft EOSs with $\Gamma=5/3$, and $3/2$. We note that for higher angular velocity, i.e., when entering the gray zone, there is no solution. This gives us the Roche limit for different EOSs. As we already noted, the maximum angular velocity in terms of $\bar\Omega$ increases for softer EOSs. Indeed as the EOS gets softer, the core of the star gets more compact, and hence, the star becomes less susceptible to tidal disruption. 

The meaning of the other curves in Figs.~\ref{Reson} and \ref{Reson2} will be described in Sec.~\ref{Res}.

\section{Perturbative expansion}
\label{Per}


\subsection{Framework}
We then consider a star in a slightly inclined spherical orbit around Kerr BHs. As before~\cite{stoc24} we study this case using a perturbative expansion assuming that the inclination angle of the orbital plane with respect to the equatorial plane is small. 
For this, we write
\begin{equation}
    b_{ia}=b_{0,ia}+\varepsilon \beta_{ia}+O(\varepsilon^2),
\end{equation}
where $b_{0,ia}$ is the equilibrium solution for equatorial circular orbits and $\beta_{ia}$ the perturbation. $\varepsilon$ is our small parameter for the expansion (see Appendix~\ref{geodesic}).

To facilitate the subsequent computations, we assume that $\alpha$ stays constant and equal to $(b_1b_2b_3)^{1/3}$. Thus, we have at first order 
\begin{equation}
    \hat b_{ia}=\alpha^{-1}(b_{0,ia}+\varepsilon \beta_{ia})=\hat b_{0,ia}+\varepsilon\hat \beta_{ia}.
\end{equation}
Then by expanding Eq. \eqref{eq32} and rescaling it in the same way as seen in the conformal correspondence theorem, we obtain the order 1 equation 
\begin{eqnarray}
    \label{eq47}
    &&\tau_*^2\alpha^3\ddot{\hat \beta}_{ia}\hat b_{0,ja}+2\tau_*\alpha^{3/2}\hat w_{ik}\dot{\hat\beta}_{ka}\hat b_{0,ja} \nonumber \\ &&
    +\hat w_{ik}\hat w_{kl}(\hat\beta_{la}\hat b_{0,ja}+\hat b_{0,la}\hat\beta_{ja}) \nonumber \\ &&
    +\frac{3}{2}\left(d\tilde A_{ik}(\hat\beta_{la})\hat b_{0,kb}\hat b_{0,jb}  
    +\tilde A_{ik}(\hat\beta_{ka}\hat b_{0,ja}+\hat b_{0,ka}\hat\beta_{ja})\right)
    \nonumber \\ &&    =-\hat C_{0,il}(\hat\beta_{la}\hat b_{0,ja}+\hat b_{0,la}\hat\beta_{ja})-\hat C_{1,il}\hat b_{0,la}\hat b_{0,ja}+\delta \hat P\delta_{ij}, 
    \nonumber \\
\end{eqnarray}
where $dA_{ik}(\hat\beta_{la})$ is the differential of $A_{ij}$ and a linear function in its argument $\hat\beta_{la}$. The linear function $dA_{ik}$ is computed using the deformation matrix of the unperturbed star $\hat b_{0,ia}$. The exact expression of $dA_{ik}(\hat\beta_{la})$ is given in Appendix \ref{moreForm}. $\hat C_{0,ij}$ and $\hat C_{1,ij}$ are the zeroth- and first-order forms of the rescaled tidal tensor.

As of now, we have 9 equations for 10 unknowns, the nine components of $\hat\beta_{ia}$ and $\delta\hat P$. Contrary to the incompressible case, we cannot assume that the star has a fixed volume. We thus need one more equation.  For this, 
we assume 
an adiabatic transformation from the equilibrium state to the perturbed state. Thus,  
\begin{equation}
    \frac{\delta P}{P}=\Gamma\frac{\delta\rho}{\rho},
\end{equation}
and we have at first order 
\begin{equation}
    \delta P =-\Gamma P\hat b_{0,ia}^{-1}\hat\beta_{ia}.
\end{equation}
Using this, we can compute the difference of the integral $\Pi$ between the equilibrium and perturbed states as 
\begin{eqnarray}
\label{deltaPressure}
    \varepsilon\delta\Pi&=&\Pi_1-\Pi_0 \nonumber \\&=&\int_{V_\varepsilon}(P(r')+\varepsilon\delta P(r'))d^3r'-\int_{V_0}P(r)d^3r \nonumber\\
    &=&-\varepsilon(\Gamma-1)\int_{V_0}P(r)\hat b_{0,ia}^{-1}\hat\beta_{ia}d^3r,
\end{eqnarray} 
where $V_\varepsilon$ designated the volume enclosed by the perturbed star and $V_0$ the equilibrium one.
Note that this last equation can be seen to be equivalent to the conservation of volume \begin{eqnarray}
\label{consVolume}
    \hat b_{0,ia}^{-1}\hat\beta_{ia}=0,
\end{eqnarray} when $\Gamma\to\infty$. This corresponds indeed to the incompressible case.

We expect to be able to split the equations in two parts: one will describe the normal oscillations of the star, and the other the feedback of the perturbed geodesics and the lower order perturbations of the star. The equations can then be seen as an excitation process of the latter term on the former one.

\subsection{Case of an inclined orbit - order 1}
\label{order1}

Since $\varepsilon$ varies in time (see Appendix~\ref{geodesic}), we assume the form of $\varepsilon(\tau)=\varepsilon_0e^{i\omega_\theta\tau}$, where $\varepsilon_0$ is the maximum precessing angle of the perturbed orbit and only the real part in $\epsilon$ has the physically relevant value. Thus, we can also write $\hat C_{1,ij}(\tau)=\hat C_{1,ij}^0e^{i\omega_\theta\tau}$, where 
\begin{eqnarray}
    \hat C_{1,13}^0=3a\varepsilon_0\frac{m\sqrt{r^2+K_0}(r^2+5K_0)}{\sqrt{K_0}r^6},
\end{eqnarray} 
with $K_0$ the Carter constant of the equatorial orbit. 
As the time varying period of $\tilde C_{1,ij}$ is fixed as $2\pi/\omega_\theta$, which we rescale as $\omega_\theta=\alpha^{-3/2}\tau_*^{-1}\hat\omega_\theta$, Eq.~\eqref{eq47} is written as
\begin{eqnarray}
    \label{eq53}
    &&-\hat\omega_\theta^2\hat \beta_{ia}\hat b_{0,ja}+2i\hat\omega_\theta\hat w_{ik}\hat\beta_{ka}\hat b_{0,ja} \nonumber\\ &&   +\hat w_{ik}\hat w_{kl}(\hat\beta_{la}\hat b_{0,ja}+\hat b_{0,la}\hat\beta_{ja}) \nonumber \\ &&    +\frac{3}{2}\left(d\tilde A_{iklm}(\hat\beta_{ma}\hat b_{0,la}+\hat b_{0,ma}\hat\beta_{la})\hat b_{0,kb}\hat b_{0,jb} \right.
    \nonumber \\ && \left.~~~~
    +\tilde A_{il}(\hat\beta_{la}\hat b_{0,ja}+\hat b_{0,la}\hat\beta_{ja})\right)
    \nonumber \\ &&
    =-\hat C_{0,il}(\hat\beta_{la}\hat b_{0,ja}+\hat b_{0,la}\hat\beta_{ja})-\hat C_{1,il}\hat b_{0,la}\hat b_{0,ja}+\delta \hat P\delta_{ij}. \nonumber \\
\end{eqnarray}

Thanks to the rescaling, we obtain similar equations as in the incompressible case. In particular, as $\hat C_{1,ij}^0$ only contains off-diagonal terms, we only need to look at the equations for the off-diagonal terms with $i\not=j$, which are the same as the ones for the incompressible model. We thus expect the first-order resonance to be present as in \cite{stoc24}, although the resonant orbital distance may differ from those in the incompressible star.
 
If we introduce $\lambda$ such that $\omega_{\theta}^2=(1+\lambda)\Omega_0^2$ where $\lambda$ is a purely relativistic correction (see Eq.~(\ref{omegat})), we have $\tilde C_{11}=(-2-\lambda)\Omega_0^2$, $\tilde C_{22}=\Omega_0^2$, and $\tilde C_{33}=(1+\lambda)\Omega_0^2$. Then, similarly to the result in \cite{stoc24}, we find the first-order resonance when
\begin{equation}
    \alpha^{3/2}\tau_*\Omega^2=\frac{3\mathcal{B}_{13}\lambda}{(3+\lambda)(1+\lambda)},
\end{equation}
where $\mathcal{B}_{13}$ is defined in Eq.~(\ref{calBdef}). 

Next, we derive a formulation for analyzing the growth rate of the resonant motion. We notice that Eq.~\eqref{eq47} can be written as linear ordinary differential equations of the first order. For this, we introduce $(e_{ij}, f_{ij})$ such that $f_{ij}=\tau_*\alpha^{3/2}\dot e_{ij}$ and $e_{ij}=\hat \beta_{ia}\hat b_{0,ja}$. Eq.~\eqref{eq47} then becomes 
\begin{eqnarray}
\label{eq55}
    &&\tau_*\alpha^{3/2}\dot{f}_{ij}+2\hat w_{ik}f_{kj} 
    +\hat w_{ik}\hat w_{kl}(e_{lj}+e_{jl}) \nonumber \\ &&
    +\frac{3}{2}\left(d\tilde A_{iklm}(e_{lm}+e_{ml})\hat b_{0,kb}\hat b_{0,jb}  
    +\tilde A_{ik}(e_{kj}+e_{jk})\right)
    \nonumber \\ &&    =-\hat C_{0,il}(e_{lj}+e_{jl})-\hat C_{1,il}\hat b_{0,la}\hat b_{0,ja}+\delta \hat P\delta_{ij}.~~~
\end{eqnarray} 

If we introduce a state vector $E_\alpha$ which encodes all the variables of the system $(e_{ij}, f_{ij}, \delta P)$, the preceding equation can be schematically written as 
\begin{equation}
\tau_*\alpha^{3/2}\dot E_\alpha+M_{\alpha\beta}E_\beta=\mathbf{C}_\alpha, 
\end{equation}
where $\mathbf{C}_\alpha$ encodes the tidal perturbation. From this equation, we can see that the eigenvalues of the matrix $M_{\alpha\beta}$ correspond to the normal modes of the star. 

At the resonance, we have $i\hat\omega_\theta E_\alpha+M_{\alpha\beta}E_\beta=0$. We then need to assume another ansatz for the time dependence of the perturbation as
\begin{eqnarray}
    E_{\alpha}&=&\mu\rho_\alpha\tau e^{i\hat\omega_\theta \tau}+\nu_\alpha e^{i\omega_\theta\tau},
\end{eqnarray} where $\mu, \rho_\alpha$, and $\nu_\alpha$ are constants, and $\mu$ denotes the growth rate of the resonant motion. While at some point the first term will be large enough that our linear analysis is not valid anymore, this ansatz allows us to compute the growth rate as the star is at or near the resonance. 

The equations satisfied by these constants can be decomposed into two systems of equations, one depending on time as $\tau e^{i\omega_\theta\tau}$, and the other as $e^{i\omega_\theta\tau}$. The one depending as $\tau e^{i\omega_\theta\tau}$ reflects the resonance condition, i.e., one the normal modes of the star $\omega_n$ coincide with $\omega_\theta$, and involves only $\rho_\alpha$
\begin{eqnarray}
    \label{eq57}
    i\hat\omega_\theta\rho_\alpha+M_{\alpha\beta}\rho_\beta=0 .
\end{eqnarray}
This implies that $\rho_\alpha$ is an eigenvector of the matrix $M_{\alpha\beta}$, which we will assume to be normalized to one. Then we find that $\mu$ can be interpreted as the growth rate of the perturbation at the resonance. 

By replacing $\omega_\theta$ with any $\omega$, one can find the normal modes $\omega_n$ with their associated eigenvectors $E_{\alpha}^n$. From this one can deduce the existence of $P_{\alpha\beta}$ a projection operator such that $P_{\alpha\beta}\rho_{\beta}=\rho_\alpha$ and $P_{\alpha\beta}E_\beta^n=0$ for any value of $n$.

One can then write $\nu_\alpha=\sigma_0 \rho_{\alpha}+\sum_n\sigma_nE_{\alpha}^n$. The equations for $\nu_{\alpha}$ are 
\begin{eqnarray}    \mu\rho_\alpha+i\hat\omega_\theta\nu_\alpha+M_{\alpha\beta}\nu_\beta=\mathbf{C}_\alpha.
\end{eqnarray}
Then by applying the projection operator $P_{\alpha\beta}$, we obtain the equation \begin{eqnarray}
\label{growthrate}
    \mu\rho_\alpha=P_{\alpha\beta}\mathbf{C}_\beta,
\end{eqnarray} 
from which we can obtain the growth rate of the perturbation $\mu$ at the resonance. $\sigma_n$ can then be determined from the initial conditions. 

Using similar notations, we can also study the growth rate of a perturbation when the star is nearby the resonance. In this case, there is a normal mode $\hat\omega_0$ in our system whose value is close to $\hat\omega_\theta$. If we assume that the response to the tidal disruption is mainly dominated by oscillations of frequency $\hat\omega_0$ and $\hat\omega_\theta$, we can assume that $E_\alpha=\mu_0\rho_{0,\alpha}e^{i\hat\omega_0\tau}+\mu_\theta\rho_{0,\alpha}e^{i\hat\omega_\theta\tau}$, where $\rho_{0,\alpha}$ is the eigenvector of $M_{\alpha\beta}$ with the eigenvalue $i\hat\omega_0$. As before, we can construct the projection operator $P_{\alpha\beta}$ associated with $\rho_{0,\alpha}$.
We must then have $i(\hat\omega_\theta-\hat\omega_0)\mu_\theta\rho_{0,\alpha}=P_{\alpha\beta}\mathbf{C}_\beta$. If we assume that our initial conditions are $E_\alpha(\tau=0)=0$, we have $\mu_\theta=-\mu_0$. Thus for $\tau\ll 1$, near the resonance, we obtain $E_\alpha \approx P_{\alpha\beta}\mathbf{C}_\beta\tau e^{i\hat\omega_\theta\tau}$. 

\subsection{Order 2}

In the previous paper, we also found resonances at the second order of perturbation in $\varepsilon$~\cite{stoc24}. To identify this, we extend our perturbation equations by one order further. 

For the second-order equations, two types of contribution, which did not appear at the first order, are present. First, one needs to add second-order contributions to the rotation rate $w_{ij}$. For this, we write 
\begin{eqnarray}
    w_{ij}=w_{0,ij}+\varepsilon^2w_{2,ij}+O(\varepsilon^4).
\end{eqnarray} $\hat w_{2,ij}$ like $\hat w_{0,ij}$ is a rotation around the axis 3. 
Second, quadratic terms in $\hat\beta_{ia}$ are necessary. However, it is soon found that these terms are not relevant for the condition of the resonances. 

We write the deformation matrix up to the second order as
\begin{eqnarray}
    b_{ia}=b_{0,ia}+\varepsilon\beta_{ia}+\varepsilon^2\gamma_{ia} +O(\varepsilon^3).
\end{eqnarray}
As before, we rescale them by $\alpha$ as
\begin{eqnarray}
    \hat b_{ia}=\hat b_{0,ia}+\varepsilon\hat\beta_{ia}+\varepsilon^2\hat\gamma_{ia}+O(\varepsilon^3).
\end{eqnarray}
Then the basic equations are written in the form
\begin{eqnarray}
\label{2ndOrderEq}
    &&\tau_*^2\alpha^3(\ddot{\hat \gamma}_{ia}\hat b_{0,ja}+\ddot{\hat\beta}_{ia} \hat\beta_{ja}) \nonumber \\ &&
    +2\tau_*\alpha^{3/2}\hat w_{0,ik}(\dot{\hat\gamma}_{ka}\hat b_{0,ja}+\dot{\hat\beta}_{ka}\hat\beta_{ja})\nonumber\\+&&\hat w_{0,ik}\hat w_{0,kl}(\hat\gamma_{la}\hat b_{0,ja}+\hat b_{0,la}\hat\gamma_{ja}+\hat\beta_{la}\hat\beta_{ja})\nonumber\\+&&\frac{3}{2}\big(d\tilde A_{iklm}(\hat\gamma_{ma}\hat b_{0,la}+\hat b_{0,ma}\hat\gamma_{la}+\hat\beta_{la}\hat\beta_{ma})\hat b_{0,kb}\hat b_{0,jb}\nonumber\\+&&\tilde A_{il}(\hat\gamma_{la}\hat b_{0,ja}+\hat b_{0,la}\hat\gamma_{ja}+\hat\beta_{la}\hat\beta_{ja})\nonumber\\+&&d^2\tilde A_{iklmno}(\hat\beta_{ma}\hat b_{0,la}+\hat b_{0,ma}\hat\beta_{la})\nonumber \\ &&\times (\hat\beta_{nb}\hat b_{0,ob}+\hat b_{0,nb}\hat\beta_{ob})\hat b_{0,kc}\hat b_{0,jc}\nonumber\\+&&d\tilde A_{iklm}(\hat\beta_{ma}\hat b_{0,la}+\hat b_{0,ma}\hat\beta_{la})\nonumber \\ && \times(\hat\beta_{ib}\hat b_{0,kb}+\hat b_{0,kb}\hat\beta_{ib})\big)\nonumber\\=&&\hat C_{0,ik}(\hat\gamma_{ka}\hat b_{0,ja}+\hat b_{0,ka}\hat\gamma_{ja}+\hat\beta_{ka}\hat\beta_{ja})
    \nonumber \\ +&&\hat C_{1,ik}(\hat\beta_{ka}\hat b_{0,ja}+\hat b_{0,ka}\hat\beta_{ja})+\hat C_{2,ik}\hat b_{0,ka}\hat b_{0,ja}\nonumber\\+&&\delta_2\hat P\delta_{ij}-\dot{\hat w}_{2,ik}\hat b_{0,ka}\hat b_{0,ja}-2\hat w_{2,ik}\hat w_{0,kl}\hat b_{0,la}\hat b_{0,ja}.
\end{eqnarray}
$d^2A_{iklmno}$ is the second derivative of $A_{ij}$ computed from the unperturbed state. The explicit form for it is given in Appendix \ref{moreForm}. 


For the second-order change of the pressure $\delta_2\hat P$, we proceed as before \begin{eqnarray}
     \delta_2\hat P=&&(\Gamma-1)\hat P\Big(-\hat b_{0,ia}^{-1}\hat\gamma_{ia}-\frac{1}{2}(\hat b_{0,ia}^{-1}\hat\beta_{ia})^2\nonumber\\&&~~~~~+\frac{1}{2}\hat b_{0,ia}^{-1}\hat\beta_{ja}\hat b_{0,jb}^{-1}\hat\beta_{ib}+\Gamma(\hat b_{0,ia}^{-1}\hat\beta_{ia})^2\Big).
 \end{eqnarray}
In the incompressible limit, $\Gamma\to\infty$, we have \begin{eqnarray}
    &&-\hat b_{0,ia}^{-1}\hat\gamma_{ia}-\frac{1}{2}(\hat b_{0,ia}^{-1}\hat\beta_{ia})^2  \nonumber\\ 
    &&+\frac{1}{2}\hat b_{0,ia}^{-1}\hat\beta_{ja}\hat b_{0,jb}^{-1}\hat\beta_{ib}+\Gamma(\hat b_{0,ia}^{-1}\hat\beta_{ia})^2=0.
\end{eqnarray} 
Here, 
$\Gamma(\hat b_{0,ia}^{-1}\hat\beta_{ia})^2$ vanishes in the incompressible limit. Indeed, from Eq.~\eqref{deltaPressure}, $\Gamma\hat b_{0,ia}^{-1}\hat\beta_{ia}=-\delta P/P$, and $\hat b_{0,ia}^{-1}\hat\beta_{ia}\to0$ as $\Gamma\to\infty$, as established in Eq.~\eqref{consVolume}, and thus, $\Gamma(\hat b_{0,ia}^{-1}\hat\beta_{ia})^2\to0$. The equation then becomes
\begin{eqnarray}
    \hat b_{0,ia}^{-1}\hat\gamma_{ia}=\frac{1}{2}\hat b_{0,ia}^{-1}\hat\beta_{ja}\hat b_{0,jb}^{-1}\hat\beta_{ib}.
\end{eqnarray}

Each source term can be decomposed in two parts as in our previous paper~\cite{stoc24}, depending on its time dependence. The first component is constant in time, and the second one varies as $e^{2i\omega_\theta\tau}$.
One then has to solve the equations for each of those time dependencies. 
However, we only want to find the resonant distances. As seen in the previous paper, we thus need to solve the equations for the time dependence in $e^{2i\omega_\theta\tau}$. Then by looking at the contributions linear in $\hat\gamma_{ia}$, one should recast the equations in matrix form and identify the orbital separation at which its determinant vanishes.
   
\subsection{Correspondence between affine equations and virial equations}

In order to clarify the continuity between the previous paper and current work, we now make the link between the current formalism and the virial equations clearer.


In the incompressible hypothesis, our starting point is an ellipsoid. We then assume that a displacement is written in the form of $x_i\to x_i+\xi_{ij}x_j$, where $\xi_{ij}$ denotes the displacement tensor. This is equivalent to performing a linear transformation $\mathcal{L}_{ij}$ on our equilibrium state: $x_i\to\mathcal{L}_{ij}x_j$, with $\mathcal{L}_{ij}=\delta_{ij}+\xi_{ij}$.

In the affine model, on the other hand, every variable is defined in relation to the spherical reference state. We apply a linear transformation from this reference state. Our perturbative expansion then concerns the transformation $b_{ia}$. In particular at first order, we apply the transformation $b_{ia}=b_{0,ia}+\varepsilon\beta_{ia}=(\delta_{ij}+\varepsilon\beta_{ib}b_{0,bj}^{-1})b_{0,ja}$. This is equivalent to applying a transformation: $\tilde{\mathcal{L}}_{ij}=(\delta_{ij}+\varepsilon\beta_{ib}b_{0,bj}^{-1})$ to the equilibrium state which is defined using $b_{0,ia}$. 

By comparing the two formalisms, we thus deduce that the displacement of the incompressible formalism can be written as 
\begin{eqnarray}
    \xi_{ij}=\beta_{ib}b_{0,bj}^{-1}.
\end{eqnarray} In the previous paper~\cite{stoc24}, we mainly used the variables $V_{i;j}^{(1)}=m_{\text{star}}\xi_{ij}b_j^2/5$. Its correspondence in the affine formalism is then given as \begin{eqnarray}
    V^{(1)}_{i;j}=\frac{m_{\text{star}}}{5}\beta_{ib}b_{0,bj}.
\end{eqnarray}
This can be also extended to higher orders, in particular for the second order as \begin{eqnarray}
    V^{(2)}_{i;j}=\frac{m_{\text{star}}}{5}\gamma_{ib}b_{0,bj}.
\end{eqnarray}

\section{Results}
\label{Res}

\subsection{Resonance orbital distance}
\label{ResOrbDistance}
\begin{figure}[t]
    \centering
         \includegraphics[width=0.5\textwidth]{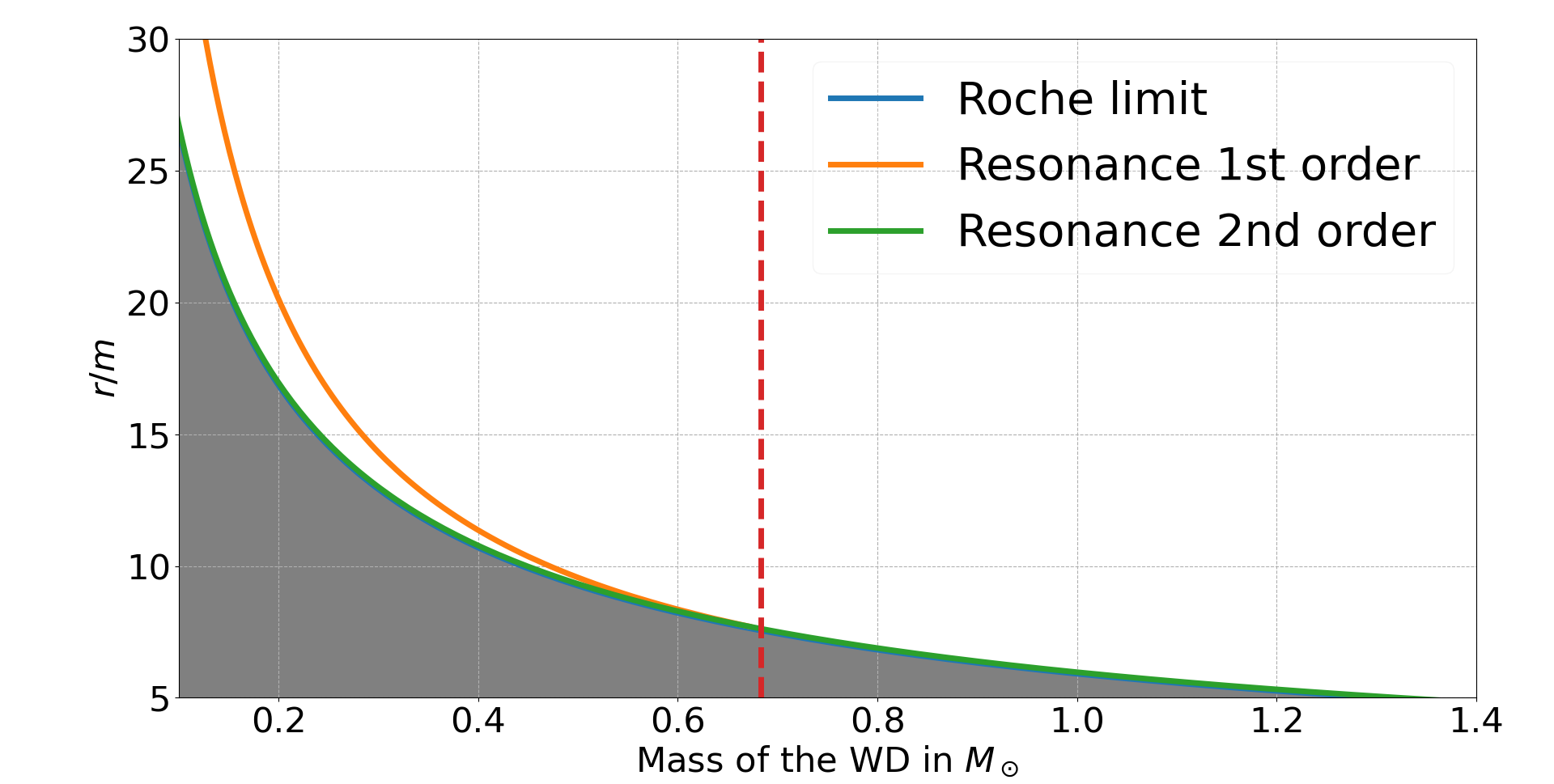}
    \caption{\label{LowDensWD} We plot the resonant distance for low density WD as a function of its mass, orbiting a $m=10^5M_\odot$ BH. We assume $\Gamma=5/3$ for the EOS of low density WD. We detail its EOS more in Sec.~\ref{ResOrbDistance}. The grey zone corresponds to the separation for which the star is tidally disrupted. We further add a red dashed line to separate the masses, below which the star first hits the first-order resonance. 
    }
\end{figure}

In our previous paper~\cite{stoc24}, we showed that when the frequency of a normal mode of an incompressible star coincides with the geodesic oscillation, a resonance is triggered.
As the overall equations have not essentially changed, it is unsurprising to find the same resonances even in compressible stars. There are, however, a couple of noteworthy changes, which will be reported in this section. 

Specifically, we here consider a white dwarf of mass $m_\mathrm{WD}$ in precessing spherical orbits around a Kerr BH of mass $m=10^5 M_{\odot} \gg m_\mathrm{WD}$ and spin parameter $a/m=0.8$. We compute the resonances for polytropic EOSs with indices of $n=1$, 3/2, and 2 ($\Gamma=2, 5/3$, and $3/2$, respectively). We note that the resonance, which we will show below, appears irrespective of the BH spin.

In Figs.~\ref{Reson} and \ref{Reson2} we plot $\tilde\Omega^2=\Omega^2/(\pi\overline\rho)$ of the first- and second-order resonances (orange and green curves, respectively) in addition to that of the Roche limit as functions of the separation of the star $r/m$. 
Here, using $\tilde\Omega$ allows us to compare the result found in the affine model with the ones from the incompressible model. 
In the following we consider the case of $m_\mathrm{WD}=0.6M_\odot$ and the white dwarf radius $R=10^4$\,km. 

For both types of resonances, for a given value of $r/m$, we observe that the value of $\tilde\Omega$ for the resonances increases in the polytropic EOS compared to the incompressible case. Moreover, with the increase of the polytropic index (decrease of the adiabatic index $\Gamma$), $\tilde\Omega$ at the resonances increases, approaching the curve of the Roche limit. In the case of the second-order resonance, it even disappears for $n=2$ ($\Gamma=3/2$). This implies that this resonance is absent for high-mass white dwarfs for which the EOS should be soft with $n \agt 2$ \cite{1939isss.book.....C}.  

For the larger values of $n$ (smaller values of $\Gamma$), the core of the star becomes more compact. Thus as we noted when we discussed the Roche limit, the star is less easily tidally disrupted (for given values of $m_\mathrm{WD}$ and $R$), and this fact can be seen in the increase of $\tilde\Omega$ at the Roche limit as the EOS becomes softer ($n$ becomes larger). 
This change in the phenomenology of the stars is also reflected in the resonances. 
In particular, we observe that white dwarfs, which may be approximated by a polytrope of $n=3/2$--2 for mass $\alt 1M_\odot$, are tidally disrupted only at a very close orbital distance to the BH. Likewise, the resonances are triggered closer to the BH. 


The first-order resonance comes from the coincidence of the geodesic oscillation with a mode driven by the Coriolis force~\cite{chan69}. It is thus a r-mode resonance. Both of them have at every point a value of the same order as $\Omega$. As the perturbative equations were the same for the polytropic and incompressible stars, we could indeed find it again. For different stellar structures, we might have slightly different resonant frequencies, and thus the resonance might be triggered at a different separation. However as the mechanism of this resonance relies mainly on the Coriolis force, we should always observe this resonance unless the EOS is not very soft. 

As we saw in Sec.~\ref{equatPlane}, a less compact white dwarf will reach one of the resonances at a more distant orbit. As long as the white dwarf is less than or as compact as our reference case with $m_\mathrm{WD}=0.6M_\odot$ and $R=10^4$\,km, then it will hit the first-order resonance. However if the white dwarf is more compact, it may hit the second-order resonance instead. 

We highlight this behavior in Fig.~\ref{LowDensWD}, where we plot the resonant distance for relatively low-mass (low density) white dwarfs as a function of its mass. We still assume that the central BH has a mass $m=10^5M_\odot$. Since our focus in particular on relatively low-mass white dwarfs, we take $\Gamma=5/3$. For this case the radius and mass of white dwarfs are approximately written in terms of the central density $\rho_\mathrm{c}$ as (e.g., \cite{shap83}) \begin{eqnarray}
    R=1.122\times10^4\left(\frac{\rho_c}{10^6 \,\mathrm{g\,cm^{-3}}}\right)^{-1/6} \text{km}, \\
    m_\mathrm{WD}=0.4964\left(\frac{\rho_c}{10^6 \,\mathrm{g\,cm^{-3}}}\right)^{1/2}M_\odot.
\end{eqnarray} 
Figure~\ref{LowDensWD} shows that for white dwarfs of mass $m_\mathrm{WD}\gtrsim 0.68M_\odot$, there will not be a first-order resonance. Thus for such massive white dwarfs, there would only be a second-order resonance, triggered close to the Roche limit. In these cases, we might observe a fairly typical tidal disruption.

Before closing this subsection, it is worth noting that r-mode resonances have already been described for relativistic binary systems~\cite{flan07, pois20}. However, these come from a gravitomagnetic tidal field, whereas the one we presented comes from a gravitoelectric field. In the case of a tidally locked star, the gravitomagnetic force is of $O(m^2R^2/r^5)$, whereas the off-diagonal gravitoelectric force is of $O(maR/r^4)$. Therefore for rapidly spinning BHs, the gravitomagnetic force is subdominant.

\subsection{Growth rate of the resonant motion in the first-order resonance}

\begin{figure}[t]
    \centering
         \includegraphics[width=0.5\textwidth]{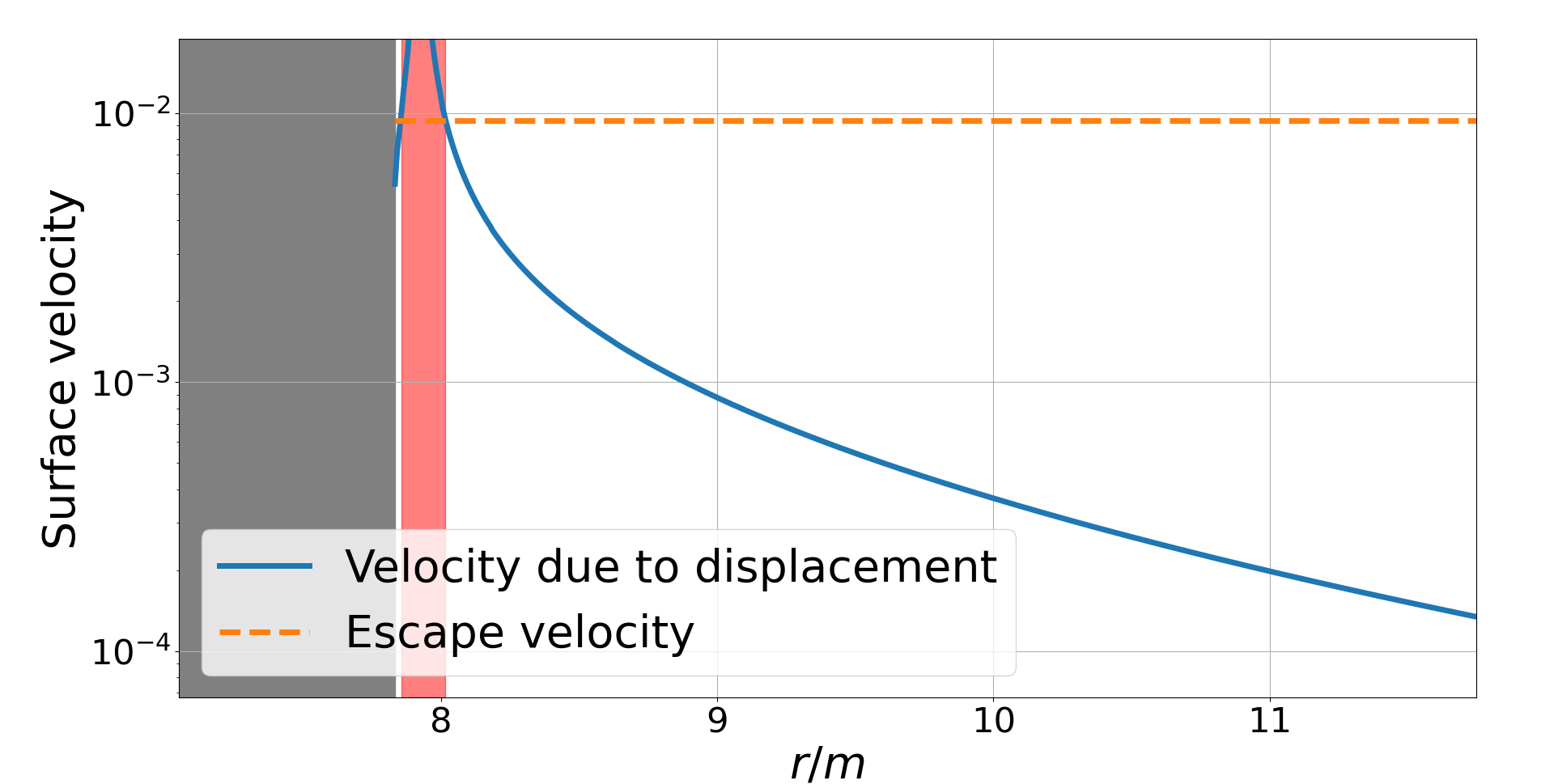}
    \caption{\label{AmpVelocity} The maximum velocity at the surface of the star due to the first-order response to the tidal perturbation for a white dwarf of $m_{\text{WD}}=0.6 M_\odot$, $R=10^4$\,km, and with a polytropic index $n=3/2$ ($\Gamma=5/3$), as a function of the separation $r/m$, where $m=10^5 M_\odot$ and $a=0.8m$. The amplitude of the perturbation on the inclination is given as $\varepsilon_0=0.5$. The dashed line shows the escape velocity for this star $\sqrt{m_{\text{WD}}/R}.$ The gray zone corresponds to the separation for which the star is tidally disrupted. We further highlight the red zone, the resonance window, where the maximum speed induced by the tidal perturbation is of the order of the escape velocity. 
    }
\end{figure}

In Fig.~\ref{AmpVelocity}, we plot the maximum velocity induced by the BH tidal force for a white dwarf of $m_{\text{WD}}=0.6 M_\odot$ and $R=10^4$\,km with $\Gamma=5/3$ at different separation $r/m$ of the BH, of $m=10^5 M_\odot$ and $a=0.8m$. We assume that the amplitude of the perturbation of the inclination is $\varepsilon_0=0.5$. The maximum velocity induced by the perturbation is computed by first solving for the variable $e_{ij}$ in Eq.~\eqref{eq55}. We can then compute the associated displacement tensor $\xi_{ij}=\beta_{ib}b_{0,bj}^{-1}=e_{ik}\hat b_{0,bk}^{-1}\hat b_{0,bj}^{-1}$ from the first-order tidal perturbation. This displacement tensor induces a velocity field $u_i=i\omega_\theta\xi_{ij}x_j$ which takes its maximum value at the surface. Here we plot $u_{\max}=R\omega_\theta\max_{ij}(\xi_{ij})$.

We compare the resulting velocity with the escape velocity at the surface of the star which is of the order $\sqrt{m_{\text{WD}}/R}$. When the induced perturbation is of the same order as this limit, we know that the linear approximation breaks down: In reality, at this stage, the growth of the perturbation would be saturated. This allows us to identify the resonant window of the separations $r$ for which the perturbation becomes large. We observe that the resonance window is narrow as $\Delta r \sim 0.15m$ (see the red region of Fig.~\ref{AmpVelocity}). 

We expect that in real systems, the perturbation will grow until the maximum velocity reaches the escape velocity. Then, a fraction of the stellar matter will be ejected from the white dwarf. By the mass loss, the radius of the white dwarf should increase with time gradually; the compactness of the white dwarf decreases. As long as the orbit of the white dwarf remains in the resonant window, this process is likely to continue. Because the orbital separation is close to that of tidal disruption, this process may trigger the eventual tidal disruption. 

During the inspiral of a white dwarf, gravitational waves are emitted. The frequency in the late inspiral phase is in the band of Laser Interferometer Space Antenna (LISA) for $m\sim 10^5M_\odot$~\cite{2017arXiv170200786A}, for which the frequency of gravitational waves from a white dwarf in circular orbits at $r/m=8$ is $\approx 30$\,mHz. When the white dwarf enters into the resonance window, it no longer behaves as a point mass, and thus, the gravitational-wave phase would be modified. After the mass ejection and subsequent tidal disruption, the gravitational-wave signal should shut down. One noteworthy point is that this shutdown occurs at an orbital frequency slightly lower than expected for a tidal disruption on the original configuration of the white dwarf. 

As some stellar matter is ejected from the white dwarf, the resonance should produce some electromagnetic signal. It is however unclear if it could be distinguished from one of a tidal disruption. Due to the special nature of the displacement involved, the signal could be modulated by the frequency of the precession $\omega_\theta$. As this frequency is not exactly equal to $\Omega$ the orbital period, one could in principle differentiate the cases where a resonance was triggered.

\begin{figure}
    \centering
         \includegraphics[width=0.5\textwidth]{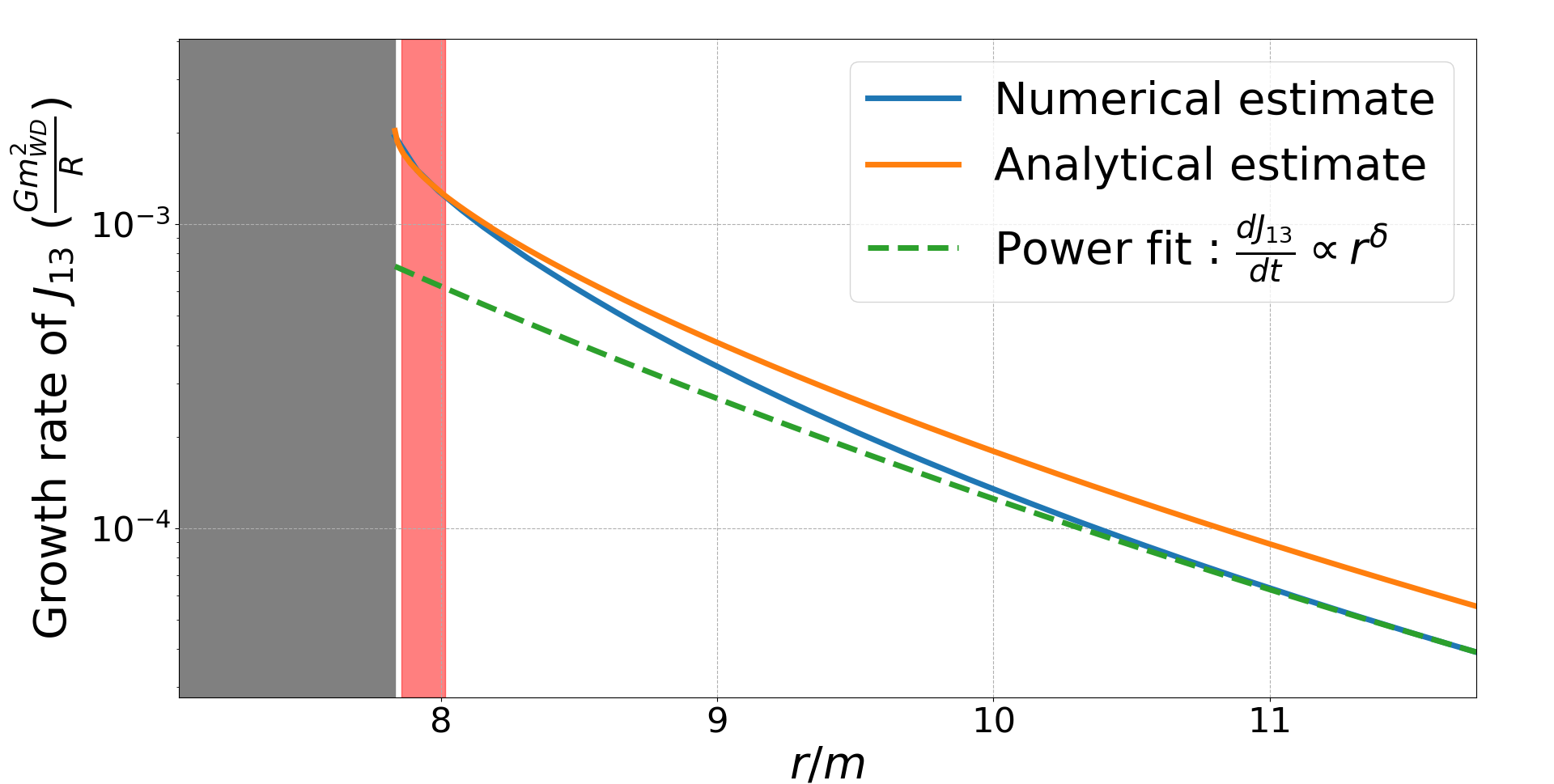}
    \caption{\label{GrowthRateFigure} Two estimates of the growth rate of the angular momentum in the $(x_1,x_3)$-plane due to the tidal perturbation as a function of the separation $r/m$. The first comes from the numerical integration of Eq.~\eqref{eq55}, and the second one is derived using the analytical formula given in Eq.~\eqref{growthrate}. We also fit the numerical estimate of the growth rate at larger separation to a power law $dJ_{13}/dt\propto r^\delta$, with $\delta\approx-7.2$. As before, we assume a white dwarf of $m_{\text{WD}}=0.6 M_\odot$, $R=10^4$\,km, and with a polytropic index $n=3/2$ ($\Gamma=5/3$) and the BH of $m=10^5 M_\odot$ and $a=0.8m$, together with the amplitude of the perturbation on the inclination $\varepsilon_0=0.5$. The gray zone corresponds to the separation for which the star is tidally disrupted and the red zone is the resonance window highlighted in Fig.~\ref{AmpVelocity}.   
    }
\end{figure}


In Fig.~\ref{GrowthRateFigure}, we plot two estimates of the linear growth rate of the angular momentum in the $(x_1,x_3)$-plane $J_{13}$, for the same system as in Fig.~\ref{AmpVelocity} assuming an initial state vector $E_\alpha(\tau=0)=0$. The first comes from the numerical integration of Eq.~\eqref{eq55}. The second is an analytical estimate of the growth rate using the formula given in Eq.~\eqref{growthrate}. We observe a good match between these two quantities in the resonance window. This is to be expected, as Eq.~\eqref{growthrate} was derived assuming that we were close to the resonance. We also notice that at the larger separation the growth rate can be fitted as a power law $\displaystyle\frac{dJ_{13}}{dt}\propto r^\delta$, with $\delta\approx-7.2$, close to $7$. This power agrees with the order-of-estimate at large radii (note that $\tilde C_{13} \propto r^{-4}$ (see Appendix~\ref{tidalTens}) and $I_{11}-I_{33} \propto r^{-3}$ for $r \gg m$ where $I_{ij}$ denotes the quadrupole moment of the star).

Let us make an estimate of the number of orbits needed for the displacement to have an important impact on the white dwarf. In the system considered, the resonance is at roughly $r/m\approx 8$. Thus $\Omega\approx \displaystyle \frac{1}{m}\left(\frac{r}{m}\right)^{-3/2}\approx 4\times 10^{-7} M_\odot^{-1}$. 
In units $c=G=1$, we have a growth rate $dJ/dt\approx 10^{-7} M_\odot c^2$. We can very roughly approximate $J\approx Iu_{\max}/R$, with $u_{\max}$ the maximum velocity typically occurring at the surface. We have in this case $I/R\approx 8\times 10^{2} M_\odot^2$. Thus after one period, we have $u=\displaystyle \frac{R}{2\pi I\Omega}\frac{dJ}{dt}\approx5\times10^{-4} c$. As the 
escape velocity $u$ can be roughly estimated as $u\approx\sqrt{m_{\text{WD}}/R}\approx10^{-2} c$, this indicates that the star needs on the order of 10--100 orbits before being seriously perturbed by the tidal forces. 

Studies of tidal disruption events have often assumed parabolic encounters of a white dwarf around a black hole \cite{teje17, gaft19, magu20}. Let us assume that our white dwarf is in a parabolic orbit around the black hole. To treat this case rigorously we would need another mathematical approach. However as the white dwarf only gets close to the black hole once, it is probable that the resonance will not have time to be triggered. Indeed from our previous order of magnitude computation, we saw that the white dwarf needs multiple orbits before the induced displacement disrupts it. 



\section{Conclusion}

In this paper, we extend our previous study~\cite{stoc24} to the case of a compressible star using the polytropic EOS. We applied the affine model formulated in \cite{cart85} and found that compressible stars could exhibit the same resonances but only in restricted cases, i.e., only for the relatively stiff EOSs. 

As remarked in \cite{stoc24}, the same analysis can be performed for weakly eccentric orbits. The perturbative equations we introduced could also be used in this case. The principles leading to the resonances can then be expected to still apply. Thus we may also expect stars in slightly eccentric orbits to have a resonance before they encounter the Roche limit, if the EOS is not too soft. 

This could be of relevance for the products of three-body interaction or a tidal capture event. Indeed, we should expect such systems to be eccentric and have a non-equatorial orbit. Thus we might observe an even wider array of resonances in such cases. 


When stars are at the resonance, or near it, the induced displacement becomes very large. The perturbative approach then breaks down. To explore it further, we need to go beyond a linear analysis. Doing this will allow us to study the impact of the resonance on the evolution of the system. It will also enable us to explore possible signatures of these resonances. For this one needs to perform numerical hydrodynamics simulations of the system. This is the subject of a follow-up work (Lam et al., in preparation).

Furthermore it has been widely shown that the actual result of a tidal resonance on an object depends on the inner structure of the object. Thus while we studied relatively simple stars, further insights could be gained on this resonance by taking more complicated inner structures into consideration.

\begin{acknowledgments}

We thank Alexis Reboul-Salze, Cédric Jockel and Alan Tsz-Lok Lam for useful discussions.
This work was in part supported by Grant-in-Aid for Scientific Research (grant No.~23H04900) of Japanese MEXT/JSPS.

\end{acknowledgments}

\appendix

\section{Spacetime metric and geodesic}
\label{geodesic}

The metric of a Kerr BH in the Boyer-Lindquist coordinates is written as
\begin{multline}
    ds^2=-\frac{\Delta}{\Sigma}(dt-a\sin^2\theta d\phi)^2+\\\frac{\sin^2\theta}{\Sigma}[(r^2+a^2)d\varphi-adt]^2+\frac{\Sigma}{\Delta}dr^2+\Sigma d\theta^2,
\end{multline} where $\Delta=r^2-2mr+a^2$ and $\Sigma=r^2+a^2\cos^2\theta$.

The time-like geodesics on the Kerr spacetime are described by the four constants of motion,  $E$: specific energy, $L$: specific angular momentum for the BH spin direction, $K$: the Carter constant~\cite{cart68}, and the mass of the test particle. The resultant geodesic equations are written as~(e.g.,~\cite{bard72})
\beqn
{dt \over d\tau}&=&{[(r^2+a^2)^2-\Delta a^2 \sin^2\theta] E - 2m r a L
  \over \Delta \Sigma},~~~\\
\Sigma^2\left({dr \over d\tau}\right)^2&=&
\left\{E(r^2+a^2)-aL\right\}^2
-\Delta (r^2 +K) \nonumber \\
&:=&\cR(r),\label{Req}\\
\Sigma^2 \left({d\theta \over d\tau}\right)^2&=&
K-a^2\cos^2\theta-{(aE \sin^2\theta-L)^2 \over \sin^2\theta}, 
\label{Thetaeq}\\
{d\varphi \over d\tau}&=&{1 \over \Delta}
          \left[{2mr aE \over \Sigma}+\left(1-{2mr \over \Sigma}\right)
            {L \over \sin^2\theta}\right],
\eeqn
where $\tau$ is an affine parameter of the geodesics. In this paper, we assume that the star has a spherical orbit with a fixed value of $r=r_0$; $\cR=0=d\cR/dr=0$ at $r=r_0$. These relations give us the two relations among $E$, $L$, and $K$.

We assume a slightly precessing orbit, i.e., $\varepsilon=\theta-\pi/2 \ll 1$. $\varepsilon$ then is a solution of the following equation  
\begin{equation}
    r^4\left(\frac{d\varepsilon}{d\tau}\right )^2=\mathcal{C}-\varepsilon^2\left[a^2(1-E_0^2)+L_0^2\right],
\end{equation} 
where $\mathcal{C}=K-(L-aE)^2$, which vanishes for equatorial orbits. We took into account the terms at $O(\varepsilon^2)$. Note that $E_0$ and $L_0$ denote $E$ and $L$ for equatorial circular orbits while for $\cC$ we need the second-order quantities in $\varepsilon$.

Then, for $r=r_0$, we obtain
\begin{equation}
\varepsilon(\tau)=\varepsilon_0\cos(\omega_{\theta}\tau), 
\end{equation} 
where 
\begin{eqnarray}
\omega_{\theta}&=&\frac{\sqrt{L_0^2+a^2(1-E_0^2)}}{r_0^2}\nonumber \\
&=&\sqrt{{r_0^2-4a \sqrt{mr_0}+3a^2 \over P_0}}\Omega_0, \label{omegat}\\
\varepsilon_0&=&\frac{\sqrt{\mathcal{C}}}{\omega_\theta}. 
\end{eqnarray}
Here $\varepsilon_0$ and $\sqrt{\mathcal{C}}$ are first-order parameters which are used for the perturbative expansion. It is worth noting that $\cos\theta(\tau)=-\varepsilon(\tau)$ at the first order.

\section{Tidal tensor}
\label{tidalTens}

The expression of the tidal tensor can be simplified by changing the frame we are working on. In a rotating frame of rotation $\vec\Omega$ along the 3-axis with magnitude $\Omega=d\Psi/d\tau$, the tidal tensor becomes~\cite{marc83}
\begin{eqnarray}
    \tilde C_{11}&=&\Big(1-3\frac{ST(r^2-a^2\cos^2\theta)}{K\Sigma^2}\Big)I_a\nonumber \\
    &&+6ar\cos\theta\frac{ST}{K\Sigma^2}I_b, \\
    \tilde C_{22}&=&I_a,\\
    \tilde C_{33}&=&\Big(1+3\frac{r^2T^2-a^2\cos^2\theta S^2}{K\Sigma^2}\Big)I_a\nonumber \\
    &&-6ar\cos\theta\frac{ST}{K\Sigma^2}I_b,\\
    \tilde C_{12}&=&0,\\ 
    \tilde C_{13}&=&3\Big[-ar\cos\theta(S+T)I_a\nonumber \\
    &&+(a^2\cos^2\theta S-r^2T)I_b\Big]\frac{\sqrt{ST}}{K\Sigma^2},\\
    \tilde C_{23}&=&0,
\end{eqnarray} 
where 
\begin{eqnarray}
I_a&=&\frac{mr}{\Sigma^3}(r^2-3a^2\cos^2\theta), \\
I_b&=&\frac{ma\cos\theta}{\Sigma^3}(3r^2-a^2\cos^2\theta), 
\end{eqnarray}
$S=r^2+K$, $T=K-a^2\cos^2\theta$, and $\Psi$ is a time-dependent angle obeying the following equation
\begin{equation}
    {d\Psi \over d\tau}=\frac{\sqrt{K}}{\Sigma}\left(\frac{E(r^2+a^2)-aL}{r^2+K}+a\frac{L-aE\sin^2\theta}{K-a^2\cos^2\theta}\right).
\end{equation}

\section{Gravitational self-potential}
\label{moreForm}


We define here the quantities necessary to evaluate the gravitational self-potential. This also establishes further the links between the affine and incompressible models. In this section, we do not make use of the Einstein summation convention.

The formula of the gravitational self-potential, $\tilde A_{ij}$, was given by Eq.~\eqref{eq22}.
%
At zeroth order, the matrix $\hat b_{ia}$ is diagonal. In terms of its components, $(\hat b_1, \hat b_2, \hat b_3)$, we have 
\begin{eqnarray}
    A_{ij}^{(0)}&=&\hat b_1\hat b_2\hat b_3\int_0^\infty\frac{\delta_{ij}du}{\hat D(\hat b_i^2+u)}
    =\mathcal{A}_{i}\delta_{ij},
\end{eqnarray} 
where $\hat D=\sqrt{(\hat b_1^2+u)(\hat b_2^2+u)(\hat b_3^2+u)}$ and 
\begin{eqnarray}
    \mathcal{A}_{i}=\hat b_1\hat b_2\hat b_3\int_0^\infty\frac{du}{\hat D(\hat b_i^2+u)}.
\end{eqnarray}
In a similar fashion, we define 
\begin{eqnarray}
    \mathcal{A}_{ij}&=&\hat b_1\hat b_2\hat b_3 \int_0^\infty\frac{du}{\hat D(\hat b_i^2+u)(\hat b_j^2+u)}. 
\end{eqnarray}

At first order, we have $S_{ij}=\sum_a [\hat b_{0,ia}\hat b_{0,ja}+\varepsilon(\hat\beta_{ia}\hat b_{0,ja}+\hat b_{0,ia}\hat\beta_{ja})]+O(\varepsilon^2)$.
Thus when only focusing on first order terms, we have
\begin{eqnarray}
    \sqrt{\det(\mathbf{S})}&=&\hat b_1\hat b_2\hat b_3\left(1+\varepsilon\sum_{a=1}^3\hat\beta_{ia}\hat b_{0,ia}^{-1}\right), \nonumber\\
    \sqrt{\det(\mathbf{S}+u\mathbf{1})}^{-1}&=&\hat D^{-1}\Big(1 \nonumber\\ 
    &-&\frac{\varepsilon}{2}(\hat b_0^2+u\mathbf{1})^{-1}_{ij}\sum_{a=1}^3(\hat\beta_{ia}\hat b_{0,ja}+\hat b_{0,ia}\hat\beta_{ja})\Big), \nonumber \\
    (\mathbf{S}+u\mathbf{1})^{-1}_{ij}&=&(\hat b_0^2+u\mathbf{1})^{-1}_{ik}¥\Big(\delta_{kj}\nonumber\\ &-&\varepsilon\sum_{a=1}^3(\hat\beta_{ka}\hat b_{0,la}+\hat b_{0,ka}\hat\beta_{la})(\hat b_0^2+u\mathbf{1})^{-1}_{lj}\Big). \nonumber\\
\end{eqnarray}
Therefore 
\begin{eqnarray}
&&    d\tilde A_{ij}(\hat\beta_{ka})=\tilde A_{ij}^{(0)}\sum_{b=1}^3\hat\beta_{ib}\hat b_{0,ib}^{-1}
    -\hat b_1\hat b_2\hat b_3\nonumber \\&\times&\sum_{b=1}^3\Biggl[\int_0^{\infty}\frac{(\hat\beta_{ib}\hat b_{0,jb}+\hat b_{0,ib}\hat\beta_{jb})du}{\hat D(\hat b_i^2+u)(\hat b_j^2+u)}\nonumber\\
    &&+\int_0^\infty\frac{(\hat b_0^2+u\mathbf{1})^{-1}_{ij}(\hat\beta_{ib}\hat b_{0,jb}+\hat b_{0,ib}\hat\beta_{jb})du}{2\hat D(\hat b_i^2+u)}\Biggr].
\end{eqnarray}
If we introduce $\mathcal{V}_{ij}=\sum_b(\hat\beta_{ib}\hat b_{0,jb}+\hat b_{0,ib}\hat\beta_{jb})$, we have 
\begin{eqnarray}
    d\tilde A_{ij}(\hat\beta_{ka})&=&\mathcal{A}_{i}\delta_{ij}\sum_{l=1}^3\frac{\mathcal{V}_{ll}}{2\hat b_l^2}-\mathcal{V}_{ij}\mathcal{A}_{ij}-\frac{1}{2}\delta_{ij}\sum_{l=1}^3\mathcal{V}_{ll}\mathcal{A}_{il}\nonumber\\
    &=&-\mathcal{V}_{ij}\mathcal{A}_{ij}+\frac{1}{2}\sum_{l=1}^3\mathcal{V}_{ll}\frac{\mathcal{B}_{il}}{\hat b_l^2},
\end{eqnarray} 
where
\begin{eqnarray}
\label{calBdef}
    \mathcal{B}_{ij}&=&\hat b_1\hat b_2\hat b_3
    \int_0^\infty\frac{udu}{\hat D(\hat b_i^2+u)(\hat b_j^2+u)}.
\end{eqnarray}


If we want to compare the preceding result with the equations provided by Chandrasekhar \cite{chan69}, we need to compare 
\begin{eqnarray}
&&\sum_{k=1}^3\sum_{b=1}^3\left(dA_{ik}(\hat\beta_{la})\hat b_{0,kb}\hat b_{0,jb}  
    +A_{ik}(\hat\beta_{kb}\hat b_{0,jb}+\hat b_{0,kb}\hat\beta_{jb})\right)
    \nonumber\\&=&-\mathcal{V}_{ij}\mathcal{A}_{ij}\hat b_j^2+\frac{1}{2}\sum_{l=1}^3\mathcal{V}_{ll}\frac{\mathcal{B}_{il}\hat b_i^2\delta_{ij}}{\hat b_l^2}+\mathcal{V}_{ij}\mathcal{A}_i\nonumber\\
    &=&\mathcal{V}_{ij}\mathcal{B}_{ij}+\frac{1}{2}\sum_{l=1}^3\mathcal{V}_{ll}\frac{\mathcal{B}_{il}\hat b_i^2\delta_{ij}}{\hat b_l^2}.    
\end{eqnarray}
The modification to the gravitational potential $\mathcal{M}_{ij}$ in Chandrasekhar's formalism is given as
\begin{eqnarray}
    \mathcal{M}_{ij}=\mathcal{V}_{ij}\mathcal{B}_{ij}+\frac{1}{2}\sum_{l=1}^3\mathcal{V}_{ll}\mathcal{A}_{il}\hat b_i^2\delta_{ij}.
\end{eqnarray}
The difference between the two equations comes from the term $\sum_b A_{ij}^{(0)}\hat\beta_{ib}\hat b_{0,ib}^{-1}$, which is equal to $0$ in the incompressible case, due to the conservation of volume Eq.~\eqref{consVolume}. 


We proceed likewise for the second order.
In Eq.~\eqref{2ndOrderEq}, we introduced the term \begin{eqnarray}
    \tilde A_{ij}^{(2)}&=&\sum_{l,m,n,o=1}^3\sum_{a,b=1}^3d^2\tilde A_{ijlmno}(\hat\beta_{ma}\hat b_{0,la}+\hat b_{0,ma}\hat\beta_{la})\nonumber \\ &&\times (\hat\beta_{nb}\hat b_{0,ob}+\hat b_{0,nb}\hat\beta_{ob})
\end{eqnarray}
This term can then be computed as before. We introduce the variables $\mathcal{V}_{i;j}=\sum_b\hat\beta_{ib}\hat b_{0,jb}$. We have $\mathcal{V}_{ij}=\mathcal{V}_{i;j}+\mathcal{V}_{j;i}$ and
\begin{eqnarray}    
    \tilde A_{ij}^{(2)}&=&\sum_{l=1}^3 \biggl[\mathcal{A}_{ij}\mathcal{V}_{i;l}\mathcal{V}_{j;l}\hat b_l^{-2}+\mathcal{A}_{ilj}\mathcal{V}_{il}\mathcal{V}_{lj}+\nonumber\\
    &&~~~+\frac{\mathcal{V}_{ll}\mathcal{V}_{ij}\mathcal{A}_{ijl}}{2}-\frac{\mathcal{V}_{ij}\mathcal{V}_{ll}\mathcal{A}_{ij}}{2\hat b_l^2}\nonumber\\
    &&~~~-\delta_{ij}\sum_{k=1}^3\biggl(\mathcal{A}_i\frac{\mathcal{V}_{l;k}\mathcal{V}_{k;l}}{\hat b_l^2\hat b_k^2}-\frac{\mathcal{V}_{lk}^2\mathcal{A}_{lki}}{4}\nonumber\\
    &&~~~-\frac{\mathcal{A}_{li}\mathcal{V}_{l;k}^2}{2\hat b_k^2}-\frac{\mathcal{V}_{ll}\mathcal{V}_{kk}\mathcal{A}_{il}}{4\hat b_k^2}\nonumber\\
    &&~~~+\frac{\mathcal{A}_{lki}\mathcal{V}_{ll}\mathcal{V}_{kk}}{8}+\frac{\mathcal{A}_i\mathcal{V}_{ll}\mathcal{V}_{kk}}{8\hat b_l^{2}\hat b_k^2}\biggr)\biggr],
\end{eqnarray} where 
\begin{eqnarray}
    \mathcal{A}_{ijk}&=&\hat b_1\hat b_2\hat b_3 \int_0^\infty\frac{du}{\hat D(\hat b_i^2+u)(\hat b_j^2+u)(\hat b_k^2+u)}.~~ 
\end{eqnarray}

\bibliography{ref}

\end{document}